\documentclass[amsmath,amssymb,prl,reprint]{revtex4-1}

\usepackage{graphicx}
\usepackage{dcolumn}
\usepackage{bm}
\usepackage{mathtools}
\usepackage{epstopdf}

\begin{document}
	
  \preprint{APS/PRL}
	
	\title{Scalar transport from deformed drops: the singular role of streamline topology}
	
	\author{Pavan Kumar Singeetham}
	\email{singeetham.pavan@gmail.com}
	\altaffiliation[Also at ]{Engineering Mechanics Unit, Jawaharlal Nehru Centre for Advanced Scientific Research, Bangalore-64, India}
	\author{Sumesh P. Thampi}%
	\email{sumesh@iitm.ac.in}
	\affiliation{%
		Department of Chemical Engineering, Indian Institute of Technology Madras, Chennai-36, India
	}%
	\author{Ganesh Subramanian}%
	\email{sganesh@jncasr.ac.in}
	\affiliation{%
		Engineering Mechanics Unit, Jawaharlal Nehru Centre for Advanced Scientific Research, Bangalore-64, India}
	
	\date{\today}
	
	\begin{abstract}
	We examine scalar transport from a neutrally buoyant drop, in an ambient planar extensional flow, in the limit of a dominant drop phase resistance. For this interior problem, we consider the effect of drop-deformation-induced change in streamline topology on the transport rate\,(the Nusselt number $Nu$). The importance of drop deformation is characterized by the Capillary  number\,($Ca$). For a spherical drop\,($Ca = 0$), closed streamlines lead to the ratio $Nu/Nu_0$ increasing with the Peclet number\,($Pe$), from unity to a diffusion-limited plateau value\,($\approx 4.1$); $Nu_0$ here denotes the purely diffusive rate of transport. For any finite $Ca$, the flow field consists of spiralling streamlines that densely wind around nested tori foliating the deformed drop interior. $Nu$ now increases beyond the aforementioned primary plateau, saturating in a secondary plateau that approaches $23.3$ for $Ca \rightarrow 0$, $Pe\,Ca \rightarrow \infty$, and appears independent of the drop-to-medium viscosity ratio. $Nu/Nu_0$ exhibits an analogous variation for other planar linear flows, although chaotically wandering streamlines in these cases are expected to lead to a tertiary enhancement regime.
	\end{abstract}
	
	\maketitle
	
	Transport of heat or mass in multiphase scenarios is relevant to diverse natural phenomena\,(growth of cloud condensation nuclei \cite{kinzer1951,beard1971, duguid1971}, nutrient uptake by microswimmers \cite{magar2003, guasto2012, stocker2012}) and industrial processes\,(fuel atomization in IC engines \cite{law1982}, spray drying \cite{patel2009}, liquid-liquid extraction \cite{wegener2014}, suspension polymerization\cite{vivaldo1997,brooks2010}). The underlying objective in many cases is calculation of the transport rate from a drop or a particle immersed in an ambient shearing flow. In non-dimensional terms, this amounts to determining the Nusselt\,($Nu$) or Sherwood\,($Sh$) number, the rate of heat or mass transport measured in diffusion units, as a function of the Peclet number\,($Pe$). The latter is defined as $Pe = \dot{\gamma} a^2/D$, and compares the relative magnitudes of convection and diffusion; $\dot{\gamma}$ is a characteristic shear rate, $a$ the particle or drop size and $D$ the thermal or mass diffusivity. Mass transport in liquids usually occurs in the convection dominant limit\,($Pe \gg 1$) - $Pe \sim O(10^3)$ for small molecules in sheared aqueous environments, but is several orders of magnitude larger for macromolecular solutes in viscous solvents. We show, for a drop in an ambient planar linear flow, that the deformation-induced alteration of the interior streamline topology has a singular effect on the transport rate in the convection dominant limit. For sufficiently large $Pe$, the enhancement in transport from a weakly deformed drop, relative to a spherical one, is enhanced more than five-fold for planar extension, and by nearly an order of magnitude for simple shear flow.
 
For rigid particles, transport for large $Pe$ occurs across a thin boundary layer, with $Nu$ scaling as the inverse of the boundary layer thickness. From a balance of the convection and diffusion time scales, the boundary layer thickness comes out to be $O(a Pe^{-\frac{1}{3}})$, with $Nu \propto Pe^{\frac{1}{3}}$. An analogous scenario holds for a drop when the ambient\,(continuous) phase resistance is dominant, the so-called exterior problem, with the boundary layer thickness scaling as $a Pe^{-\frac{1}{2}}$ on account of the interfacial slip, and $Nu \propto Pe^{\frac{1}{2}}$.  The transport for large $Pe$ also depends sensitively on the streamline topology\cite{ganesh2006a,ganesh2006b,deepak2018part2}, and boundary-layer-driven enhancement occurs only when the exterior flow has an open-streamline\cite{acrivos1965,leal2007,gupalo1972,gupalo1975,polyanin1984} or open-pathline\cite{banerjee2021} topology. Owing to the unbounded domain, such a topology is a common occurrence for the exterior problem, either in the Stokes limit itself\cite{deepak2018}, or due to deviations from Stokesian hydrodynamics\cite{ganesh2006c,ganesh2006a,ganesh2006b,ganesh2007,deepak2018part2}. In contrast, a closed-streamline topology is more common for a confined domain like the drop interior. While a generic three-dimensional ambient linear flow leads to chaotically wandering streamlines within a spherical drop in the Stokes limit\cite{stone1991, sabarish2021}, and a $Nu$ that increases algebraically with $Pe$\cite{bryden1999,sabarish2021}, many  canonical flows including uniform flow, linear extensional flows, and the family of planar linear flows, lead to closed interior streamlines \cite{sabarish2023}. Earlier efforts that have examined transport from a spherical drop in these flows, in the limit of a dominant drop-phase resistance\,(the interior problem), have shown that $Nu$ plateaus at a value of order unity for $Pe \rightarrow\infty$\cite{newman1931, kronig1951, Brignell1975,prakash1978, johns1966, watada1970}, implying diffusion-limited transport due to the closed-streamline topology. 

Herein, for the first time, we consider the interior problem for a deformed drop in an ambient linear flow. While there has been extensive research on drop deformation and breakup in linear flows \cite{taylor1934,rumscheidt1961,torza1971,bentley1986a,bentley1986b,rallison1984,stone1994}, substantially less attention has been paid to the correlation between deformation and streamline topology. The importance of drop deformation is characterized by the capillary number, $Ca = \mu \dot{\gamma}a/\Gamma$, $\mu$ being the ambient fluid viscosity and $\Gamma$ the coefficient of interfacial tension. While one has closed streamlines and diffusion-limited transport for $Ca = 0$\,(spherical drop), on account of Stokesian reversibility constraints\cite{newman1931, kronig1951, Brignell1975,prakash1978, johns1966, watada1970}, this must no longer be true for finite $Ca$. However, earlier experiments\cite{torza1971} and computations\cite{kennedy1994} have suggested the persistence of closed streamlines within a deformed drop in a simple shear flow. We show, using the analytical velocity field for small $Ca$, and supporting boundary integral computations, that drop deformation does open up closed streamlines. The aforementioned sensitive dependence of transport on streamline topology implies that the $Nu-Pe$ relationship is qualitatively affected by the altered streamline topology at finite $Ca$. For an ambient planar extensional flow, Langevin simulations show that a spiralling-streamline topology leads to a singularly enhanced transport for $Pe \gg 1$ for any non-zero $Ca$. In contrast, for an ambient axisymmetric extension, closed streamlines persist even at finite $Ca$ owing to symmetry constraints, and as a result, the $Nu-Pe$ relationship remains virtually unaltered for small $Ca$.
		
	Consider a neutrally buoyant Newtonian drop of viscosity $\lambda \mu$ in a Newtonian ambient undergoing a planar linear flow, $\bm{u}^\infty = \bm{\Gamma} \cdot \bm{x}$, with $\textstyle \bm{\Gamma}=2\dot{\gamma}\!\!\begin{bmatrix}
		0 & 1 & 0 \\
		\beta & 0 & 0\\
		0 & 0 & 0
	\end{bmatrix}$ being the (transpose of the)\,velocity gradient tensor. Planar linear flows are a one-parameter family with $\beta \in [-1,1]$; $\beta = -1, 0$ and $1$ correspond to solid-body rotation, simple shear and planar extension, respectively. While our primary focus is on planar extension\,($\beta = 1$), with $\bm\Gamma = \bm E$\,($\bm{E}$ being the strain-rate tensor), we discuss the streamline topology for other flows with non-unity $\beta$ towards the end of the article, along with the implications for transport. Assuming viscous effects to be dominant, the governing equations, in non-dimensional form, are:
	\begin{eqnarray}\label{eq1}
		&	\lambda \nabla^{2}\bm{\hat{u}}-\nabla \hat{p}=0, \quad \nabla \cdot \bm{\hat{u}}=0,\\ \label{eq2}
		&	\nabla^{2}\bm{u}-\nabla p=0, \quad \nabla \cdot \bm{u}=0,
	\end{eqnarray}    
with $\bm{\hat{u}}\,(\bm{u})$ and $\hat{p}\,(p)$ being the interior\,(exterior) velocity and pressure fields. Here, $a, \dot{\gamma}a$ and $\mu \dot{\gamma}$ have been used as the characteristic length, velocity and stress scales, respectively, $a$ being the radius of the undeformed spherical drop. Equations (\ref{eq1}-\ref{eq2}) are subject to the following boundary conditions:
	\begin{equation}\label{eq3}
		\left. 
		\begin{array}{ll}
			\bm{u} \rightarrow \bm{u}^{\infty} \quad \text{for} \quad \bm{x} \rightarrow \infty, \\
			\bm{u}=\bm{\hat{u}}\,(\text{Velocity continuity at the interface}),\\
\bm{u} \cdot \bm{n} = 0\,(\text{Steady drop shape}), \\
  (\bm{\sigma}\!-\!\lambda \bm{\hat{\sigma}})\!\cdot\!\bm{n} =\!\frac{1}{Ca}(\nabla \!\cdot\! \bm{n})\bm{n}\,			(\text{Interfacial stress balance}), \\
		\end{array}
		\!\!\right \}		
	\end{equation}
	Here, $\bm{n}$ is the outer unit normal to the drop surface, and $\bm{\hat{\sigma}}=-\lambda^{-1}\hat{p} \bm{I}+ (\nabla \bm{\hat{u}}+\nabla \bm{\hat{u}}^{T})$ and $\bm{\sigma}=-p \bm{I}+ (\nabla \bm{u}+\nabla \bm{u}^{T})$ are the interior and exterior stress tensors, respectively. It is straightforward to calculate the disturbance fields, and the perturbed drop shape, for small $Ca$ and $\lambda Ca \ll 1$, using a domain perturbation technique\cite{leal2007}. This has been done across several earlier efforts\cite{cox1969,barthes1973, rallison1980,vlahovska2009,leal2007,arun2012}, and the results, to $O(Ca)$, have been tabulated in the supplemental material for convenience\cite{SM}. 
 
\begin{figure}
		\centering
		\includegraphics[scale=0.435]{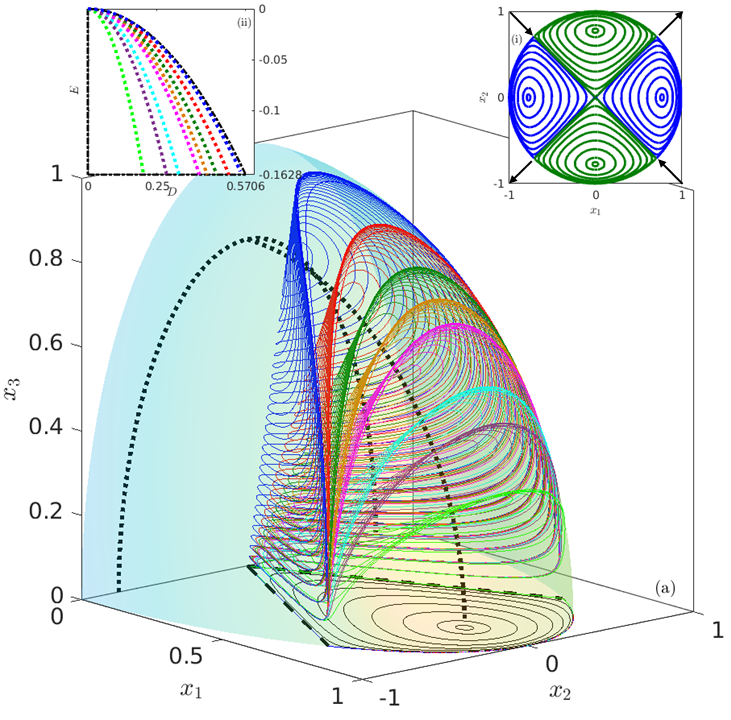}\\
		\includegraphics[scale=0.395]{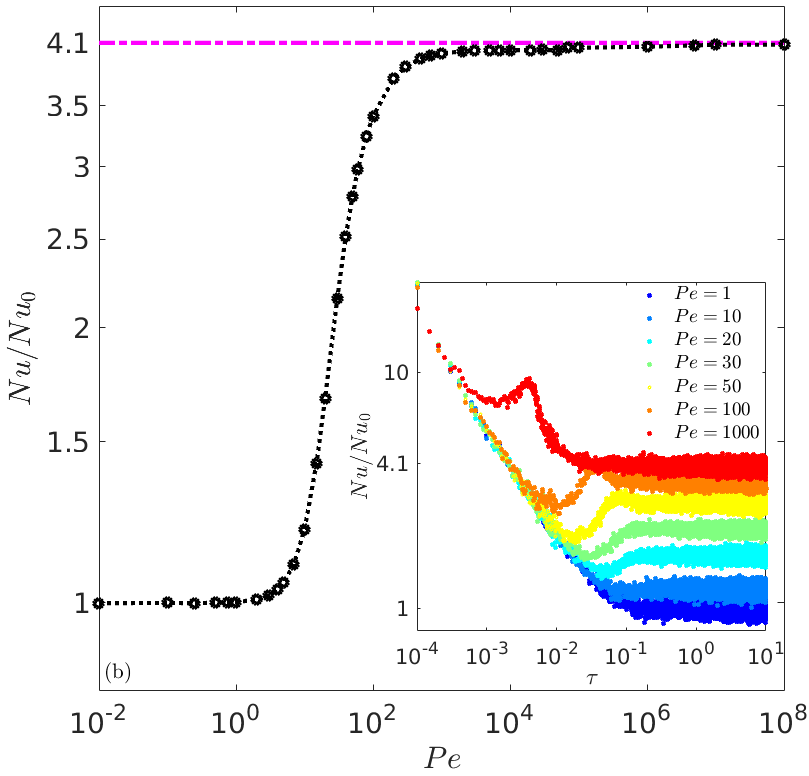}
		\caption{(a) Interior streamlines in an octant of a spherical drop\,($Ca = 0$) in an ambient planar extension; the inplane streamlines are depicted separately in inset (i); the streamlines are shown as ordered pairs on the $D-E$ plane in inset (ii). (b) $Nu/Nu_0$ v/s $Pe$ for an ambient planar extension\,($Ca = 0)$; the time dependent $Nu/Nu_0$-curves appear in an inset.}\label{fig1} 
	\end{figure}

We will use the expression for $\hat{\bm u}(\bm{x})$, to $O(Ca)$, to examine the streamline topology within the deformed drop for small but finite $Ca$. For $Ca = 0$, almost all interior streamlines are closed, and may be obtained as the curves of intersection of two one-parameter families of invariant surfaces \cite{cox1968,torza1971,powell1983,deepak2018,deepak2018part2}. Denoting the parameters as $D$ and $E$, the invariant surfaces are defined by $\displaystyle	x_{3}=D(1-r^2)^{-\frac{1}{3}}$ and $\displaystyle	x_{2}=\pm r\left( \frac{1}{2}-\frac{(1-\beta)(1+\lambda)}{2 r^2(1+\beta)}+\frac{E}{r^2(1-r^2)^{(2/3)}}\right)^{\frac{1}{2}}$, with $-0.1628 \leq E \leq 0.1628$ and $-0.5706 \leq D \leq 0.5706$; for $\beta = 1$ alone, both families are independent of $\lambda$. Each ordered pair $(D,E)$ corresponds to a closed streamline\cite{deepak2018}, the exceptions being fixed\,(stagnation) points, and streamlines connecting such points. Fig\,\ref{fig1}a depicts the closed streamlines in an octant of the spherical drop, along with the quarter-circle of fixed points\,(a black-dotted curve corresponding to $E=-0.1628, D\in[0,0.5706]$) that `threads' these streamlines, and the in-plane streamlines\,($D=0$) in the corresponding quadrant. The streamlines in other octants may be obtained by symmetry; the quarter-circle, when extended by symmetry, leads to two orthogonally oriented fixed-point circles that intersect at $(0,0,\pm\sqrt{\frac{3}{5}})$. The streamlines in all four quadrants of the flow gradient\,($x_1-x_2$) plane appear separately as inset (i) on the right; inset (ii) on the top left depicts the interior streamlines, in the main figure, as points in the $D-E$ plane.
 
The scalar transport problem is now solved by integrating the Langevin equations for the individual tracers, $d\bm{x}= Pe\, \bm{u}(\bm{x}) d\tau+ \sqrt{2}~d\bm{W}(\tau)$\cite{gardiner2004}, using a $4^{th}$-order Runge-Kutta method. Here, $\bm{u}(\bm{x})$ is the velocity field convecting the tracer at $\bm{x}(\tau)$, and $\tau$ is the non-dimensional time in units of $a^2/D$, $D$ being the scalar diffusivity. $d\bm{W}(\tau)$ is the Wiener process satisfying $\langle d\bm{W}(\tau) \rangle =0$, $\langle d\bm{W}(\tau^{\prime}) d\bm{W}(\tau) \rangle= \bm{I}|\tau'-\tau|$. We start from a uniform initial distribution of tracers corresponding to an initial concentration field $c(\bm{x},0)=1\,\forall\,|\bm{x}| < 1$. The perfect-absorber condition at the drop surface\,($c(1,\tau)=0$), pertinent to the interior problem, is implemented by removing tracers crossing the drop boundary\cite{SM}, this continual removal making the problem an inherently unsteady one. $Nu$ is defined as the normalized rate of change of the volume-averaged concentration: $\displaystyle	Nu=-\frac{2}{3\bar{c}}\frac{\partial \bar{c}}{\partial \tau}$ where $\displaystyle \bar{c}(\tau)= \frac{3}{4\pi} \oint_{|\bm{x}|\leq 1}\!\!\!c(\bm{x},\tau)d\rm{V}$\cite{kronig1951, juncu2010}. While $Nu$ is in general a function of $\tau$, for all cases examined, $c(\bm{x},\tau)$ approaches a simple exponential decay for sufficiently long times, and accordingly, $Nu$ approaches two-thirds of the largest\,(least negative) eigenvalue of the convection-diffusion operator in this limit\footnote{Note that this definition would yield the same transport rate as that obtained from the usual definition, based on the normal derivative of the concentration at the drop surface, when a true steady state is made possible due to a uniformly distributed source contribution\,(say).} 
 
Fig\,\ref{fig1}b shows $Nu/Nu_0$ in the long-time limit as a function of $Pe$, for $Ca = 0$ and $\lambda = 1$, for an ambient planar extension, with the normalized transport rates for arbitrary times appearing in an inset; $Nu_0$ here corresponds to the long-time diffusive rate of transport\,($Pe = 0$), and equals $(2/3)\pi^2$. The inset curves show a monotonic decay to a long-time plateau value for small $Pe$, reflecting the self-adjointness of the diffusion operator\,(all eigenvalues are real and negative). For large $Pe$, the decay acquires an oscillatory character, corresponding to the transient adjustment of the iso-scalar contours and streamlines on time scales of $O(\dot{\gamma}^{-1}Pe^{\frac{1}{3}})$. The main figure shows $Nu/Nu_0$ increasing from unity to a plateau value of $4.1$ for $Pe \gtrsim O(100)$. The plateau arises because tracers are rapidly convected around closed streamlines for larger $Pe$, with the long-time transport controlled by the much slower rate at which they diffuse across these streamlines\footnote{One may, in principle, derive a two-dimensional streamline-averaged diffusion equation in the coordinates $D$ and $E$, governing the scalar concentration field for $Pe \rightarrow \infty$, whose solution should yield the value of the 2$D$-diffusion-limited $Nu$-plateau; however, components of the streamline-averaged diffusivity tensor in this equation have to be determined numerically even for the simpler case of a uniform flow\cite{kronig1951}. An approximate numerical determination of, along these lines of the $Nu$-plateau, has been done for a rigid sphere enveloped by closed streamlines in an ambient simple shear\cite{acrivos1971}.}. Thus, the plateau value corresponds to the rate of diffusion across closed streamlines that are coincident with iso-scalar contours for times longer than $O(\dot{\gamma}^{-1}Pe^{\frac{1}{3}})$. The $\lambda$-independence of the streamline topology for planar extension implies that the effect of $\lambda$ may be simply accounted for via a re-scaled Peclet number, $Pe(1+\lambda)^{-1}$, the plateau value above being independent of $\lambda$. While the interior streamline topology is $\lambda$-dependent for other planar linear flows\cite{powell1983,deepak2018part2}, precluding the above scaling, almost all interior streamlines are still closed for $Ca = 0$, and $Nu$ therefore exhibits an analogous $Pe$-dependence\,(see Fig\,\ref{fig4} below). The diffusion-limited plateau equals $1.08$ for $\lambda = 1$ for simple shear\,($\beta = 0$), with $Nu$ in solid-body rotation\,($\beta =-1$) equaling unity regardless of $Pe$ or $\lambda$.

\begin{figure}
		\centering
	        \includegraphics[scale=0.415]{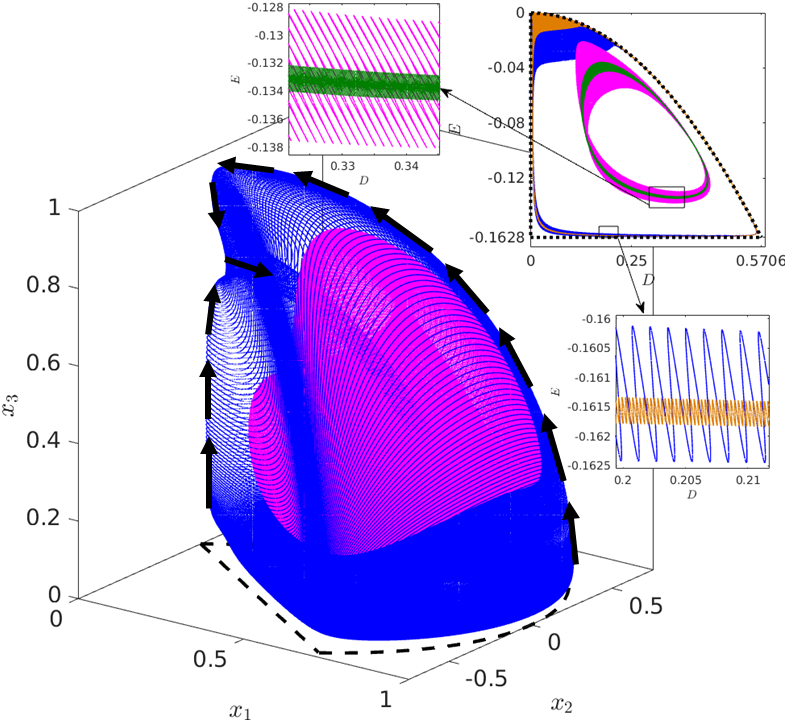}
		\caption{A pair of spiralling streamlines, within a deformed drop octant, in an ambient planar extension for $Ca=0.005, \lambda=1$: the blue and magenta curves correspond to initial points $(0.7,0,0.05)$ and $(0.5,0,0.3)$, respectively, with black arrows indicating the direction of spiralling. The top right inset shows the streamlines in the $D-E$ plane, while the smaller secondary insets highlight the two-time-scale structure and the decreasing amplitude of the fast wiggles when $Ca$ reduces from $0.005$ to $0.001$.}\label{fig2} 
	\end{figure}

Fig\,\ref{fig2} depicts a pair of streamlines within a weakly deformed drop\,($Ca$ small but finite), again in an ambient planar extension. Each streamline is seen to have a tightly spiralling character, and ends up winding densely around an invariant torus for sufficiently long times. A one-parameter family of such tori foliates the interior of the drop octant, with the limiting singular torus comprising the bounding surfaces of the deformed octant\,(that include portions of the $x_1\pm x_2=0$ planes), and a perturbed form of the quarter-circle of fixed points shown in Fig\,\ref{fig1}a. $D$ and $E$ are no longer constant along a finite-$Ca$ streamline, but are adiabatic invariants\cite{bajer1990,bajer1992} that vary on a time scale of $O(Ca^{-1}\dot{\gamma}^{-1})$, at leading order, this being much longer than the $O(\dot{\gamma}^{-1})$ period of circulation around the zero-$Ca$ closed streamlines. This two-time scale structure is illustrated via the primary inset in Fig\,\ref{fig2} where small-amplitude fast oscillations are seen to be superposed on a slow $O(Ca)$ drift. The amplitude of these oscillations decreases with decreasing $Ca$ as illustrated by the pair of smaller secondary insets\,(for $Ca =0.005$ and $0.001$). The streamlines asymptote to closed contours on the $D-E$ plane for $Ca \rightarrow 0$, for times much greater than $O(Ca^{-1}\dot{\gamma}^{-1})$. These closed curves may, in principle, be obtained from solving an autonomous system of ODE's in $D$ and $E$, derived using the method of averaging, and imply the existence of a secondary adiabatic invariant that remains constant on time scales of $O(Ca^{-1}\dot{\gamma}^{-1})$, and parameterizes these contours\footnote{An expression for this secondary invariant has been derived for the simpler problem of a translating drop\cite{bajer1990,bajer1992}.}. 
	\begin{figure}
		\centering
		\includegraphics[scale=0.375]{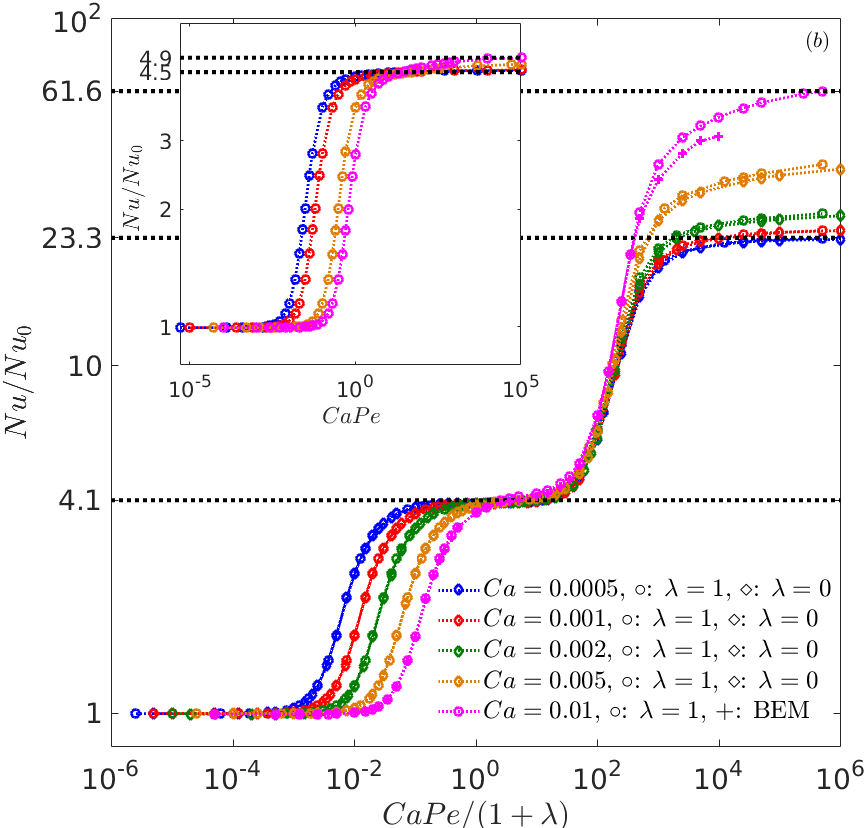}
		\caption{ $Nu/Nu_{0}$\,v\!/\!s\,$PeCa(1+\lambda)^{-1}$ for various $Ca~(=0.0005,~0.001,~0.002, ~0.005,~0.01)$. The main plot pertains to an ambient planar extension, while the inset in plots $Nu/Nu_{0}$\,v\!/\!s\,$PeCa$, for $\lambda = 1$, for an ambient biaxial axisymmetric extensional flow.}\label{fig3}
	\end{figure}
 
Similar to the inset in Fig\,\ref{fig1}b,  with increasing $Pe$, there is a transition from an overdamped to an oscillatory approach to the long-time plateau. Importantly, the plateau value is no longer bounded from above by $4.1$. Instead, as $Pe$ increases beyond $10^4$, the plateau value rises above $4.1$, eventually asymptoting to $24.5$ for $Pe = 10^8$. Fig\,\ref{fig3}b plots the long-time plateau values against $PeCa(1+\lambda)^{-1}$ for $\lambda=0$ and $1$, and for $Ca$ ranging from $0.0005$ to $0.01$\footnote{The latter value is more an order of magnitude lower than the critical value for breakup\cite{bentley1986a}, ensuring validity of the weak deformation assumption underlying use of the $O(Ca)$ velocity field.}. For all $Ca$, $Nu/Nu_0$ first increases from unity to the zero-$Ca$ plateau of $4.1$ that now serves as an intermediate asymptotic regime corresponding to $Pe \gg 1, PeCa \ll 1$. As $PeCa$ increases to unity and beyond, $Nu/Nu_0$ increases further, eventually saturating in a secondary plateau for $PeCa \rightarrow \infty$. While the primary plateau is determined by diffusion along the two coordinates orthogonal to the one along the zero-$Ca$ closed streamlines, the transport rate corresponding to the secondary plateau is determined by (one-dimensional)\,diffusion across the invariant tori like those shown in  Fig\,\ref{fig2}, and that are traced out by the finite-$Ca$ spiralling streamlines. While the secondary plateau value is in general $Ca$-dependent, increasing to $61.6$ for $Ca = 0.01$, it becomes independent of $Ca$ for sufficiently small $Ca$, approaching $23.3$ for $Ca \rightarrow 0$. Thus, remarkably, for any finite $Ca$ however small, one obtains a nearly six-fold enhancement over the primary plateau\,($23.3$\,\,v/s\,\,$4.1$) for sufficiently large $Pe\,(\gg Ca^{-1})$. 
 
 While the zero-$Ca$ velocity field for planar extension exhibits a simple $(1+\lambda)^{-1}$\!-scaling, as mentioned earlier, the $O(Ca)$ correction exhibits a more complicated dependence on $\lambda$\,(see expressions given in \cite{SM}). Nevertheless, the secondary plateau values for $\lambda = 0$ and $1$, in Fig\,\ref{fig3}b, nearly coincide for small $Ca$, suggesting that the underlying adiabatic invariant that parameterizes the small-$Ca$ tori is only weakly dependent on $\lambda$. Towards validating our results based on the $O(Ca)$ velocity field, we have used the boundary integral method\,(BEM)\cite{kennedy1994,pozrikidis1992} to compute the velocity field that is then used in the Langevin simulations described above. The BEM-cum-Langevin simulations confirm the increase in $Nu/Nu_0$ beyond the primary plateau, while also showing the disappearance of the separation between the primary and secondary plateaus with increasing $Ca$\cite{SM}.
 
In order to emphasize the critical role of streamline topology in determining the $Nu\!-\!Pe$ relationship, we have also considered a drop in an ambient biaxial axisymmetric extension. The requirement of axisymmetry implies that the finite-$Ca$ streamlines continue to be closed curves restricted to a meridional plane. The unaltered streamline topology for non-zero $Ca$ implies that the large-$Pe$ transport, even for a deformed drop, must occur via diffusion across closed finite-$Ca$ streamlines. This is confirmed by the $Nu/Nu_0$\,vs\,$Pe$ curves in the inset of Fig\,\ref{fig3}(b) which show only a single large-$Pe$ plateau even for finite $Ca$, and that exhibits a modest increase from $4.5$ for $Ca = 0$ to about $4.9$ for $Ca =0.01$ and $Pe Ca \gg 1$\,\footnote{It is worth mentioning that, unlike $Ca = 0$, the finite-$Ca$ transport is not invariant to flow reversal in the limit $Pe \rightarrow \infty$. However, on account of the perturbative effect of drop deformation, the large-$Pe$ transport rate for an ambient uniaxial extension is expected to differ only by $O(Ca)$ from the diffusion-limited plateau shown for biaxial extension.}. The significance of the disparate $Nu-Pe$ relationships for planar and axisymmetric extension, in Fig\,\ref{fig3}(b), is best appreciated when one accounts for the virtually identical character of these two flows in other scenarios. Both axisymmetric and planar extension are prototypical examples of strong flows that lead to material points separating in time at an exponential rate. Macromolecules in both flows undergo a coil-stretch transition, leading to polymer solutions exhibiting an extension-thickening rheology that often manifests as a dramatic departure from Newtonian behavior. Further, the critical $Ca$ exhibits an identical dependence on $\lambda$ for both flows, transitioning from a $\lambda^{-\frac{1}{6}}$-scaling regime for small $\lambda$ to a plateau for $\lambda \rightarrow \infty$ \cite{acrivos1978, hinch1980}\footnote{The original asymptotic analysis for the critical $Ca$ in the limit $\lambda \rightarrow 0$, for planar extension, relied on treating this flow as an ambient axisymmetric extension plus a three-dimensional perturbation \cite{hinch1980}. The results for axisymmetric extension had been derived in an earlier effort \cite{acrivos1978}, and those for planar extension were obtained for finite amplitudes of the said perturbation.}, pointing to the similar nature of drop deformation and breakup. Yet, the differing nature of the finite-$Ca$ interior streamlines leads to a profound difference in scalar transport characteristics!     

Finally, Fig\,\ref{fig4} shows that the altered streamline topology for non-zero $Ca$, and its role in transport enhancement for sufficiently large $Pe$, persists for planar linear flows other than planar extension. Inset of Fig\,\ref{fig4} shows the $Nu$\,v/s\,$Pe$ curves for $Ca = 0$, for $\lambda =1$ and $\beta = 0, 0.25, 0.5, 0.75$ and $1$, while the main Fig\,\ref{fig4} shows the corresponding curves for $Ca = 0.001$ and $0.005$. The difference between the primary and secondary plateaus, for $Ca \rightarrow 0$, is seen to be even greater than that for planar extension. For example, the secondary plateau for $Ca=0.005, \beta=0.5$ is about $12.4$, which is more than eight times the primary plateau value of $1.51$. An interesting feature of the $Nu-Pe$ curve, for $\beta = 0.75$ and $Ca = 0.005$, is the much more gradual approach towards a secondary plateau. The associated streamline pattern, for $Ca = 0$, is known to exhibit both homoclinic and heteroclinic connections, and preliminary evidence suggests that the broken connections at finite $Ca$ lead to chaotically wandering streamlines in this case. The delayed plateauing tendency might therefore be related to a tertiary enhancement regime arising from Lagrangian chaos, and will be taken up in a future investigation.	
	\begin{figure}
		\centering
            \includegraphics[scale=0.35]{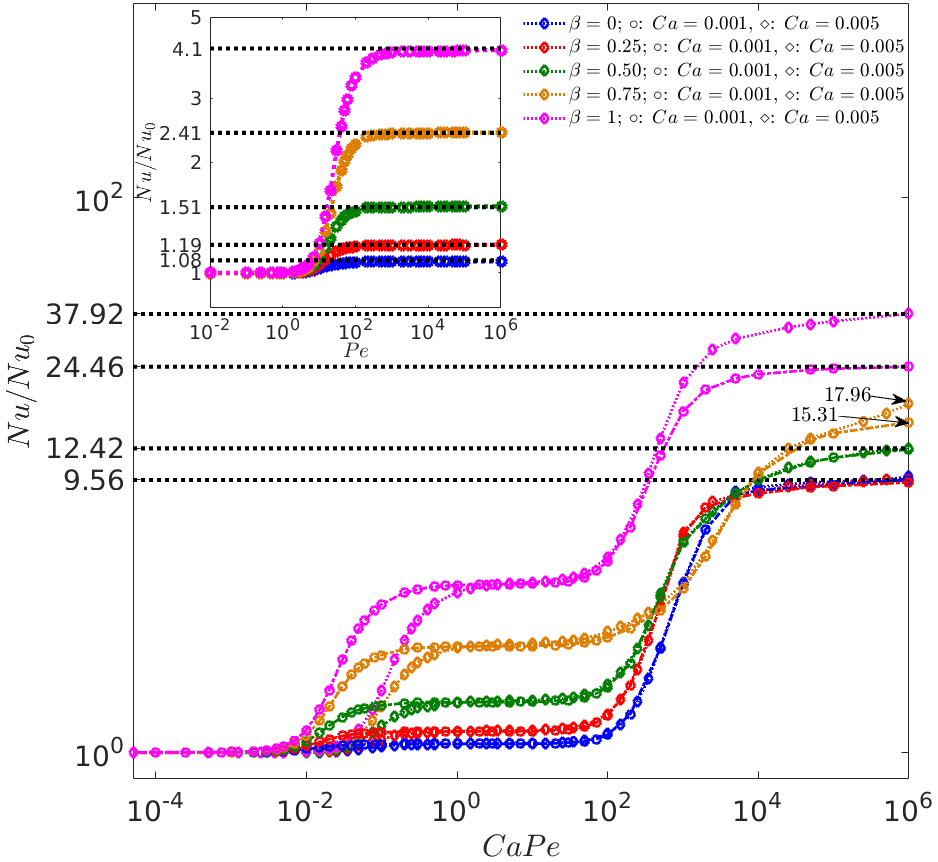}
		\caption{$Nu/Nu_{0}$\,v/s\,$CaPe$, for $Ca =0.005$, $\lambda = 1$, for a drop immersed in different members of the planar linear flow family ($\beta=0,~0.25,~0.5,~0.75,~1$), while the inset is $Nu/Nu_{0}$\,v/s\,$Pe$, for $Ca =0$.}\label{fig4} 
	\end{figure}

 \begin{acknowledgments}
Numerical computations reported here were carried out using the Param Yukti facility provided under the National Supercomputing Mission available with
JNCASR and AQUA available at IIT Madras. The authors thank the institutes for providing these facilities. P.K.S. acknowledges Akshith Patidar (Master's student at IITM) for providing the simulation data for the axisymmetric flow.  
	\end{acknowledgments}
	
	\bibliographystyle{apsrev4-1}
	\bibliography{apssamp}

\end{document}


\preprint{APS/PRL}
	
	\title{Scalar transport from deformed drops: the singular role of streamline topology}
	
	\author{Pavan Kumar Singeetham}
	\email{singeetham.pavan@gmail.com}
	\altaffiliation[Also at ]{Engineering Mechanics Unit, Jawaharlal Nehru Centre for Advanced Scientific Research, Bangalore-64, India}
	\author{Sumesh P. Thampi}%
	\email{sumesh@iitm.ac.in}
	\affiliation{Department of Chemical Engineering, Indian Institute of Technology Madras, Chennai-36, India
	}%
	\author{Ganesh Subramanian}%
	\email{sganesh@jncasr.ac.in}
	\affiliation{Engineering Mechanics Unit, Jawaharlal Nehru Centre for Advanced Scientific Research, Bangalore-64, India}
	
	\date{\today}
 \begin{abstract}
 \end{abstract}
 
	\maketitle
 \section{Supplementary Material }\label{appA}
\subsection{Velocity and pressure fields for a drop in an ambient planar linear flow}\label{appA1}	
	\begin{figure}
		\centering
		\includegraphics[scale=0.35]{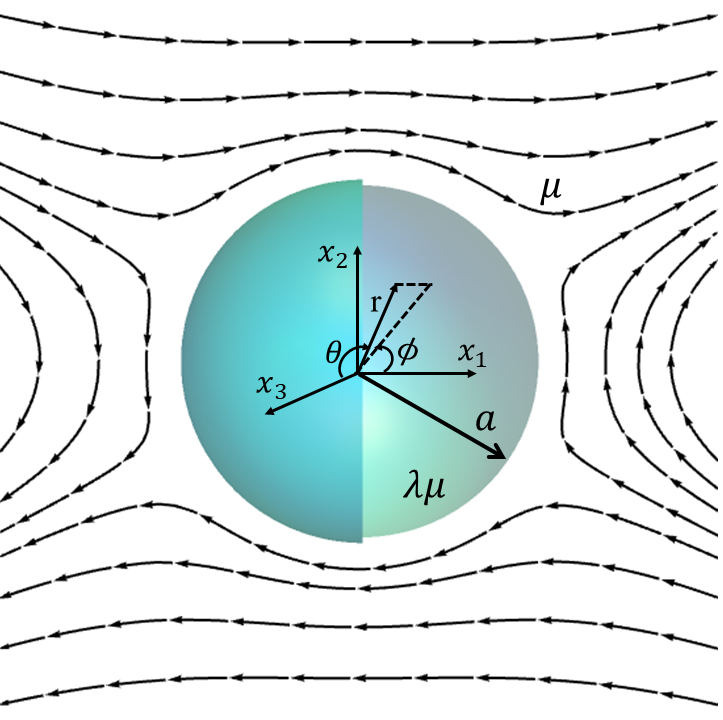}
		\caption{A schematic of a drop of radius $a$ suspended in an ambient planar linear flow.}\label{fig1a}	\end{figure}
\begin{figure}
		\centering
			\includegraphics[scale=0.135]{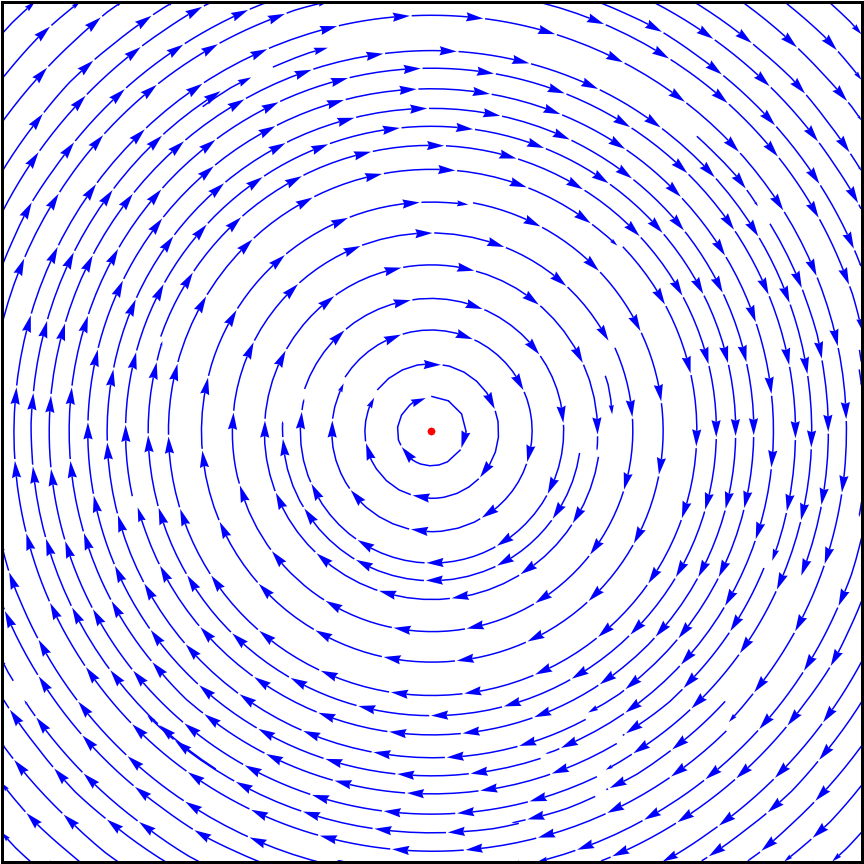}
			\label{betam1}
				\includegraphics[scale=0.135]{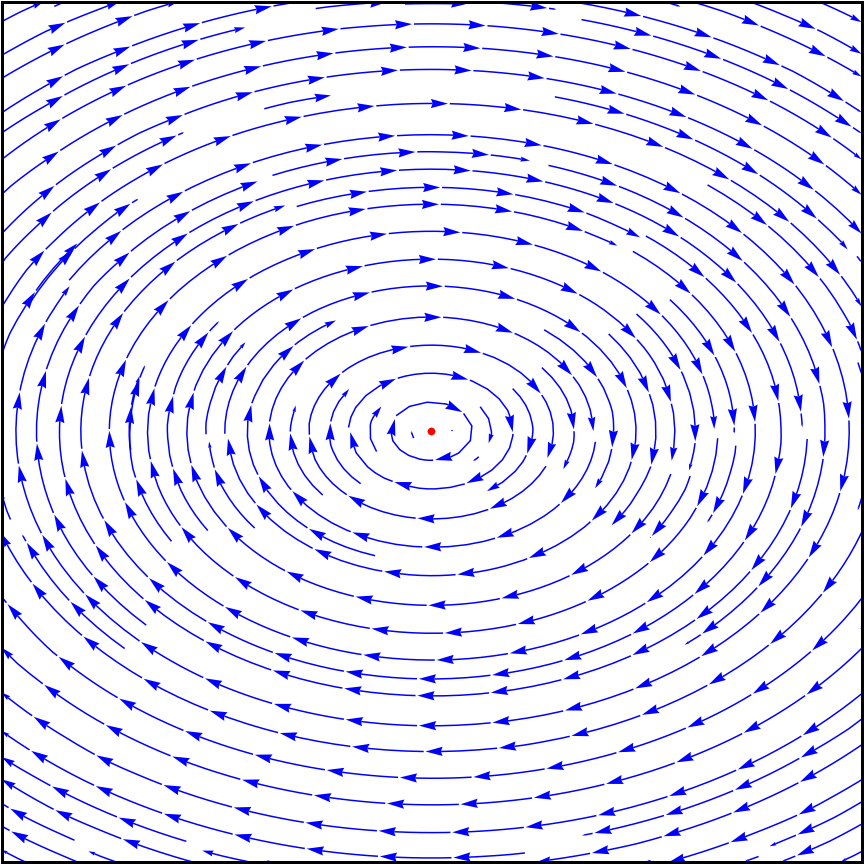}
			\label{betaless0}
			\includegraphics[scale=0.135]{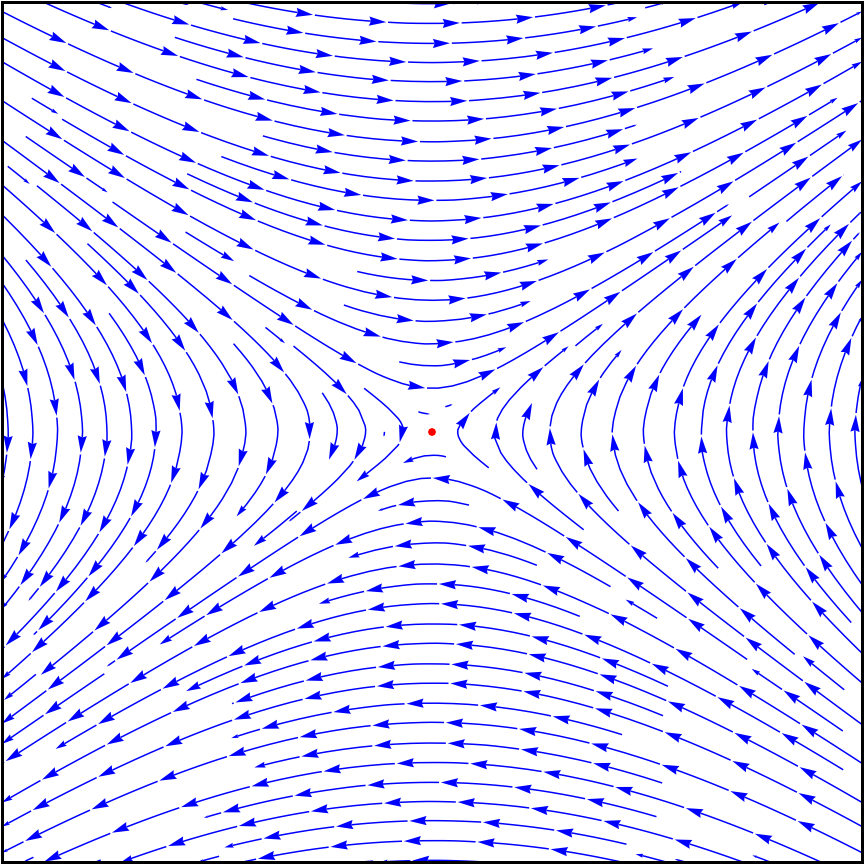}
			\label{betagrt0}
			\includegraphics[scale=0.135]{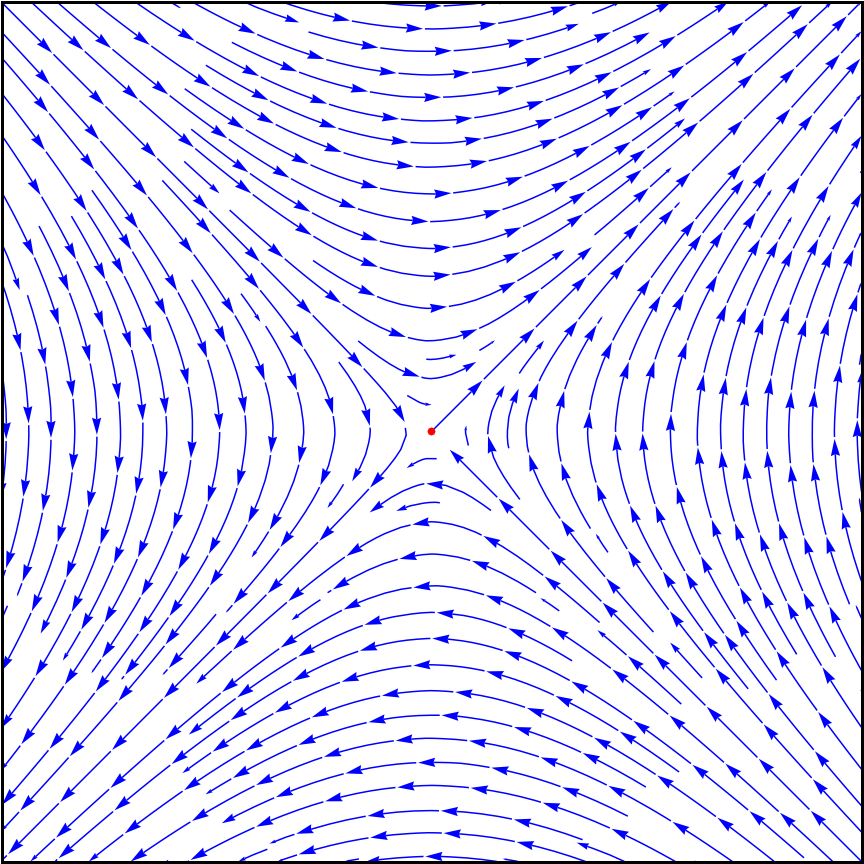}
			\label{beta1}\\
   (a) \hspace{4cm} (b) \hspace{4cm} (c) \hspace{4cm} (d)\\
		\caption{Planar linear flows from a one-parameter family, where the parameter $\beta$ ranges from $-1$ to $1$: (a) Solid-body rotation ($\beta=-1$), (b) an elliptic linear flow with $-1 \leq \beta \leq 0$, (c) a hyperbolic planar linear flow with $0 \leq \beta \leq 1$, and (d) Planar extension ($\beta=1$). }\label{fig2l}
	\end{figure}

In this section, we provide detailed expressions for both interior and exterior velocity and pressure fields induced by a neutrally buoyant drop in an ambient planar linear flow. The schematic in Fig\,\ref{fig1a} shows a nearly spherical drop of radius $a$ in an ambient hyperbolic planar linear flow. Consider a neutrally buoyant Newtonian drop of viscosity $\lambda \mu$ in a Newtonian fluid of viscosity $\mu$ that is undergoing a planar linear flow, $\bm{u}^\infty = \bm{\Gamma} \cdot \bm{x}$, with $\textstyle \bm{\Gamma}=2\dot{\gamma}\!\!\begin{bmatrix}
		0 & 1 & 0 \\
		\beta & 0 & 0\\
		0 & 0 & 0
	\end{bmatrix}$ being the (transpose of the) velocity gradient tensor. The associated rate of strain and vorticity tensors are given by $\bm{E}=\dot{\gamma}(1+\beta)\!\!\begin{bmatrix}
		0 & 1 & 0 \\
		1 & 0 & 0\\
		0 & 0 & 0
	\end{bmatrix}$ and $\bm{\Omega}=\dot{\gamma}(1-\beta)\!\!\begin{bmatrix}
		0 & 1 & 0 \\
		-1 & 0 & 0\\
		0 & 0 & 0
	\end{bmatrix}$, respectively. Planar linear flows are a one-parameter family with $\beta \in [-1,1]$; $\beta = -1, 0$ and $1$ correspond to solid-body rotation, simple shear, and planar extension, respectively. $-1 \leq \beta \leq 0$ corresponds to elliptic linear flows with closed streamlines, and $0 \leq \beta \leq 1$ corresponds to hyperbolic planar linear flows; see Fig\,\ref{fig2l}. Assuming inertial effects to be negligible, the governing Stokes equations and equations of continuity, in non-dimensional form, are:
	\begin{eqnarray}\label{eq1}
		&	\lambda \nabla^{2}\bm{\hat{u}}-\nabla \hat{p}=0; \quad \nabla \cdot \bm{\hat{u}}=0,\\ \label{eq2}
		&	\nabla^{2}\bm{u}-\nabla p=0; \quad \nabla \cdot \bm{u}=0,
	\end{eqnarray}    
with $\bm{\hat{u}}\,(\bm{u})$ and $\hat{p}\,(p)$ being the interior\,(exterior) velocity and pressure fields. Here, $a, \dot{\gamma}a$ and $\mu \dot{\gamma}$ have been used as the characteristic length, velocity, and stress scales, respectively, $a$ being the radius of the undeformed spherical drop. Equations (\ref{eq1}-\ref{eq2}) are subject to the following boundary conditions \cite{leal2007,arun2012}:
	\begin{equation}\label{eq3}
		\left. 
		\begin{array}{ll}
			\bm{u} \rightarrow \bm{u}^{\infty} \quad \text{for} \quad \bm{x} \rightarrow \infty, ~~~\bm{\hat{u}}  ~~ \text{finite at}  ~~  \bm{x}=0, \\
			\bm{u}=\bm{\hat{u}}\,(\text{Velocity continuity at the interface}),\\
\frac{1}{\left| \nabla F \right|} \frac{\partial F}{\partial t }+ Ca (\bm{u} \cdot \bm{n})=0\,(\text{Kinematic boundary condition at the interface}), \\
  (\bm{\sigma}\!-\!\lambda \bm{\hat{\sigma}})\!\cdot\!\bm{n} =\!\frac{1}{Ca}(\nabla \!\cdot\! \bm{n})\bm{n}\,			(\text{Interfacial stress balance for a clean interface}), \\
		\end{array}
		\!\!\right \}		
	\end{equation}
	Here, $\bm{n}$ is the outer unit normal to the drop surface, $\bm{\hat{\sigma}}=-\lambda^{-1}\hat{p} \bm{I}+ (\nabla \bm{\hat{u}}+\nabla \bm{\hat{u}}^{T})$ and $\bm{\sigma}=-p \bm{I}+ (\nabla \bm{u}+\nabla \bm{u}^{T})$ are the interior and exterior stress tensors, respectively, and $t$ in the kinematic condition is non-dimensionalized by the interface relaxation time scale $Ca \dot{\gamma}^{-1}$. Under the action of the imposed flow, the drop interface deforms, attaining a steady non-spherical shape for $Ca$ less than a critical value $Ca_{c}$ ($Ca_{c} \simeq 0.12$ for $\lambda=1$ and an ambient planar extension, and is higher for the other members of the planar linear flow family). For small $Ca$, one may describe the drop shape using a scalar function, $F(\bm{x}_{s},t)=r-\left(1+Ca f \left(\bm{x}_{s},t \right)\right)=0$, to $O(Ca)$, where $r = |\bm{x}|$ and $\bm{x}_{s}$ is the position of the interface, and then calculate the disturbance fields as well as the perturbed shape via a domain perturbation technique.
 
  For small $Ca$ and $\lambda Ca \ll 1$, the above calculation has been done across several earlier efforts \cite{cox1969,barthes1973, rallison1980,vlahovska2009,leal2007,arun2012}. One expands the velocity and pressure fields as $\hat{\bm{u}}=\hat{\bm{u}}^{(0)} +Ca\,\hat{\bm{u}}^{(1)}\,(\bm{u} =\bm{u}^{(0)} +Ca\,\bm{u}^{(1)})$ and $
		\hat{p}=\hat{p}^{(0)} +Ca\,\hat{p}^{(1)}\,(p=p^{(0)}+Ca\,p^{(1)})$, respectively, with the expressions for the $O(1)$ and $O(Ca)$ fields being given by:
	\begin{equation} \label{eq6}
		\begin{aligned}
			\bm{u}^{(0)}=\bm{\Omega \cdot x} +\left(1-\frac{6 c_{3}^{(0)}}{r^{5}}\right)&\bm{E\cdot x}+\left(\frac{c_{1}^{(0)}}{2r^{5}}+\frac{15 c_{3}^{(0)}}{r^{7}}\right)\bm{x \cdot (x \cdot E \cdot x)},
		\end{aligned}
	\end{equation}
 \begin{equation}\label{eq5}
		p^{(0)}=\frac{c_{1}^{(0)}}{r^{5}}(\bm{x \cdot E \cdot x}),
	\end{equation}
	\begin{equation}\label{eq8}
		\begin{aligned}
			\bm{u}^{(1)}&=\Big(\frac{c_{5}^{(1)}}{2r^{3}}+\frac{3c_{6}^{(1)}}{r^{5}}-\frac{3c_{16}^{(1)}}{r^{5}} +\frac{30 c_{23}^{(1)}}{r^{7}} \Big)\bm{x (E:E)}+\left(\frac{6c_{6}^{(1)}}{r^{5}}-\frac{6c_{16}^{(1)}}{r^{5}}+\frac{120c_{23}^{(1)}}{r^{7}} \right)\bm{(E \cdot E) \cdot x}  \\
			&-\Big( \frac{3c_{5}^{(1)}}{2r^{5}}+\frac{30c_{6}^{(1)}}{r^{7}}-\frac{15c_{16}^{(1)}}{r^{7}}+\frac{420c_{23}^{(1)}}{r^{9}}\Big)\bm{x (x \cdot (E \cdot E) \cdot x)}-\left( \frac{15 c_{6}^{(1)}}{r^{7}}+\frac{420 c_{23}^{(1)}}{r^{9}} \right)\bm{(x \cdot E) (x\cdot E \cdot x)}\\
			&+\left(\frac{105 c_{6}^{(1)}}{2r^{9}}+\frac{945 c_{23}^{(1)}}{r^{11}} \right)\bm{x (x \cdot E \cdot x)^{2}} +\left(\frac{3c_{25}^{(1)}}{r^{5}} \right)(\bm{\Omega \cdot E }) \bm{\cdot x}-\left(\frac{3c_{25}^{(1)}}{r^{5}} \right)(\bm{E \cdot \Omega} ) \bm{\cdot x}\\
			&-\left(\frac{3c_{8}^{(1)}}{2r^{5}}+\frac{15c_{25}^{(1)}}{r^{7}} \right)\bm{x (x \cdot (\Omega \cdot E) \cdot x)},
		\end{aligned}
	\end{equation}
 \begin{equation}\label{eq7}
		\begin{aligned}
			p^{(1)}&=\left(\frac{c_{5}^{(1)}}{r^{3}}+\frac{6c_{6}^{(1)}}{r^{5}}  \right)\bm{E : E}-\left( \frac{3c_{5}^{(1)}}{r^{5}}+\frac{60c_{6}^{(1)}}{r^{7}}\right)\bm{x \cdot (E \cdot E) \cdot x}+\frac{105c_{6}^{(1)}}{r^{9}}(\bm{x \cdot E \cdot x})^{2}-\frac{3c_{8}^{(1)}}{r^{5}} \bm{x \cdot (\Omega \cdot E) \cdot x},
		\end{aligned}
	\end{equation}
	\begin{equation}\label{eq11}
		\begin{aligned}
			\bm{\hat{u}}^{(0)}=&\bm{\Omega \cdot x}+ \left(d_{2}^{(0)}-6 d_{3}^{(0)}r^{2} \right) \bm{E \cdot x}+  \left( \frac{12 d_{3}^{(0)}}{5}\right)\bm{x \cdot (x \cdot E \cdot x)},
		\end{aligned}
	\end{equation}
 	\begin{equation} \label{eq10}
		\hat{p}^{(0)}=\left(-\frac{126 d_{3}^{(0)}}{5} \right)(\bm{x \cdot E \cdot x}),
	\end{equation}
	\begin{equation}\label{eq13}
		\begin{aligned}				\bm{\hat{u}}^{(1)}&=\bigg(\frac{d_{5}^{(1)}r^2}{2}+\frac{d_{7}^{(1)}}{2}+d_{15}^{(1)}-3d_{16}^{(1)} r^{2}-\frac{120 d_{23}^{(1)} r^{4}}{7} \bigg)\bm{x (E:E)}	+\Big(d_{13}^{(1)}-6d_{14}^{(1)} r^{2} -6 d_{16}^{(1)} r^{2} +120 d_{23}^{(1)} r^{4} \Big )\bm{(E \cdot E) \cdot x}\\
			&+\Big( -\frac{3 d_{5}^{(1)}}{2}+15d_{16}^{(1)}+\frac{360 d_{23}^{(1)}r^{2}}{7} \Big)\bm{x (x \cdot (E \cdot E) \cdot x)}+\bigg(15 d_{14}^{(1)} -420d_{23}^{(1)}r^{2} \bigg)\bm{(x \cdot E) (x\cdot E \cdot x)}\\
			&+120d_{23}^{(1)} \bm{x (x \cdot E \cdot x)^{2}}+\Big(d_{18}^{(1)}+3d_{25}^{(1)}r^{2} \Big)\bm{(\Omega \cdot E ) \cdot x}+\Big(d_{19}^{(1)}-3d_{25}^{(1)}r^{2} \Big)\bm{(E \cdot \Omega ) \cdot x}\\
			&-\bigg(\frac{3  d_{8}^{(1)}}{2}+15d_{25}^{(1)} \bigg)\bm{x (x \cdot (\Omega \cdot E) \cdot x)},
		\end{aligned}
	\end{equation} 
	\begin{equation}\label{eq12}
		\begin{aligned}		\hat{p}^{(1)}&=\left(d_{5}^{(1)}r^{2}+6d_{6}^{(1)}r^{4}+d_{7}^{(1)}  \right)\bm{E : E}-\Big(3d_{5}^{(1)}+60d_{6}^{(1)}r^{2} \Big)\bm{x \cdot (E \cdot E) \cdot x}+105d_{6}^{(1)} (\bm{x \cdot E \cdot x})^{2}
			-3d_{8}^{(1)} \bm{x \cdot (\Omega \cdot E) \cdot x}.
		\end{aligned}
	\end{equation}
	The deformed drop shape, to $O(Ca)$, is given by 
	\begin{equation}\label{eq14}
		r=1+Ca f^{(0)},
	\end{equation}
	where $f^{(0)}=b_{1}^{(1)}\bm{E:}\frac{\bm{xx}}{r^2}
$.	
The expressions for the unknown constants $c_{i}^{(j)}$, $d_{i}^{(j)}$ ($i=1-25$), and $b_1^{(1)}$, appearing in the above equations, are given below: 
	\begin{equation}\label{eq17}
	\left.	
  \begin{array}{ll}
			c_{1}^{(0)}=&-\frac{2 +5\lambda}{\lambda+1},~ c_{2}^{(0)}=0,~ c_{3}^{(0)}=\frac{\lambda}{6 \lambda+6},
			~d_{1}^{(0)}=\frac{21 \lambda}{2(\lambda+1)},   ~d_{2}^{(0)}=-\frac{3}{2 \lambda+2},	  ~d_{3}^{(0)}=-\frac{5 }{12 (\lambda+1)},
		\end{array}
   \right\}
	\end{equation}
	\begin{equation}\label{eq18} 
  \left.
	\begin{array}{ll}	
   c_{4}^{(1)}=0, ~c_{5}^{(1)}=\frac{64 +732 \lambda+1179 \lambda^2+475~\lambda^3}{140 (\lambda+1)^3},~c_{6}^{(1)}=-\frac{352+1138 \lambda+855 \lambda^2 }{1080 (\lambda+1)^2}, c_{8}^{(1)}=\frac{(16 +19\lambda )^2}{60 (\lambda+1)^2},~c_{9}^{(1)}=0, ~c_{12}^{(1)}=0, \\~ c_{14}^{(1)}=\frac{352 +1138 \lambda+855 \lambda^2}{1080 (\lambda+1)^2},	  ~ c_{15}^{(1)}=0, ~c_{16}^{(1)}=\frac{-1568 +458 \lambda+4915 \lambda^2+2565 \lambda^3}{7560 (\lambda+1)^3},~ c_{17}^{(1)}=0,~c_{18}^{(1)}=0,
			 ~	 c_{19}^{(1)}=0, ~	 c_{20}^{(1)}=0,\\ ~ c_{22}^{(1)}=0, ~ c_{23}=\frac{64 +796 \lambda+855\lambda^2 }{15120 (\lambda+1)^2},
			~	c_{24}^{(1)}=0,~ c_{25}^{(1)}=-\frac{32 +86 \lambda+57 \lambda^2}{120 (\lambda+1)^2}, 
   \end{array}
   \right\}
			\end{equation}
	\begin{equation}\label{eq19}
 \left.
		\begin{array}{ll}
			d_{4}^{(1)}=0, ~ 	d_{5}^{(1)}=\frac{16+83 \lambda+76 \lambda^2 }{10 (\lambda+1)^3}, ~
			d_{6}^{(1)}=\frac{11 (16 +19\lambda )}{432 (\lambda+1)^2},~
			d_{7}^{(1)}=\frac{\left(32 -554 \lambda-703 \lambda^2 \right)}{60 (\lambda+1)^2}, d_{8}=-\frac{7 \left(48 +89 \lambda+38 \lambda^2 \right)}{40 \lambda (\lambda+1)^2}, ~\\
			d_{9}^{(1)}=0,~ 
			d_{10}^{(1)}=0, ~
			d_{11}^{(1)}=0, ~
			d_{12}^{(1)}=0, ~d_{13}^{(1)}=-\frac{9 \left(16 +3 \lambda-19\lambda^2 \right)}{140 (\lambda+1)^3},~
			d_{14}^{(1)}=-\frac{2 (16 +19\lambda)}{27 (\lambda+1)^2},\\
		d_{15}^{(1)}=\frac{\left(-224 +3942 \lambda +8853 \lambda^2+4579 \lambda^3\right)}{840 (\lambda+1)^3},~	d_{16}^{(1)}=\frac{ \left(400 +1307 \lambda+988 \lambda^2\right)}{756 (\lambda+1)^3},~	d_{17}^{(1)}=0,
			~d_{18}^{(1)}=-\frac{304 \lambda^2+617 \lambda+304}{80 (\lambda+1)^2}, \\
   ~			d_{19}^{(1)}=\frac{304 \lambda^2+617 \lambda+304}{80 (\lambda+1)^2}, ~	d_{20}^{(1)}=0,  ~
			d_{21}^{(1)}=0,   ~
			d_{22}^{(1)}=0,~d_{23}^{(1)}=-\frac{7  (16 +19\lambda )}{4320 (\lambda+1)^2},~
			d_{24}^{(1)}=0, ~
			d_{25}^{(1)}=\frac{48 +89 \lambda+38 \lambda^2}{48 (\lambda+1)^2},
		\end{array}
  \right\}
	\end{equation}
 \begin{equation}\label{eq20}
     b_{1}^{(1)}=\frac{16+ 19\lambda}{8 (\lambda+1)}.
 \end{equation}
 Here, the superscripts $j=0$ and $1$ correspond to the $O(1)$ and $O(Ca)$ terms in the respective disturbance fields. While the above expressions pertain to a general linear flow, the Cartesian components that result for an ambient planar linear flow are given by:
    \begin{equation}\label{eq21}
      \begin{aligned}
   \hat{u}^{(0)}_{1}= \frac{x_{2} \left((2-2 \beta) \lambda -5 \beta-1 +(\beta +1) (x_{1}^2+5 x_{2}^2+5 x_{3}^2 )\right)}{2 (\lambda +1)},
    \end{aligned}
    \end{equation}
    \begin{equation}\label{eq22}
      \begin{aligned}
    \hat{u}^{(0)}_{2}=  \frac{x_{1} \left((2 \beta -2) \lambda -\beta -5 + (\beta +1) (5x_{1}^2+ x_{2}^2+5 x_{3}^2)\right)}{2 (\lambda +1)},
     \end{aligned}
    \end{equation}
    \begin{equation}
      \begin{aligned}\label{eq23}
  &\hat{u}^{(0)}_{3}=  -\frac{2 (\beta +1) x_{1} x_{2} x_{3}}{\lambda +1},
     \end{aligned}
    \end{equation}
\begin{equation}\label{eq24}
    \begin{aligned}
        \hat{u}^{(1)}_{1}&=-\bigg((\beta +1) (19 \lambda +16) x_{1} \Big(9 (-127 \beta +239 \lambda +139)+560 (\beta +1) (\lambda +1) x_{1}^4-x_{1}^2 \big(-378 (\beta -1) \lambda ^2\\
        &+(191 \beta +2081) \lambda +677 \beta +350 (\beta +1) (\lambda +1) x_{2}^2-910 (\beta +1) (\lambda +1) x_{3}^2+1811\big)-2870 (\beta +1) (\lambda +1) x_{2}^4\\
        &-9 x_{2}^2 \big(-\beta  (\lambda  (98 \lambda +741)+631)+\lambda  (98 \lambda -251)+280 (\beta +1) (\lambda +1) x_{3}^2-337\big)+126 (\beta -1) \lambda ^2 \big(5 x_{3}^2-8\big)\\
        &+x_{3}^2 \big(573 \beta +350 (\beta +1) x_{3}^2-1317\big)+\lambda  \big(-2259 \beta+350 (\beta +1) x_{3}^4+3 (509 \beta -541) x_{3}^2\big)\Big)\bigg)\Bigg/ (2520 (\lambda +1)^3),
    \end{aligned}
\end{equation}
\begin{equation}\label{eq25}
    \begin{aligned}
  \hat{u}^{(1)}_{2}& =  \bigg((\beta +1) (19 \lambda +16) x_{2} \big(9 (-139 \beta +251 \lambda +127)+2870 (\beta +1) (\lambda +1) x_{1}^4+x_{1}^2 \big(350 (\beta +1) (\lambda +1) x_{2}^2\\
  &+9 \big(98 (\beta -1) \lambda ^2-(251 \beta +741) \lambda -337 \beta +280 (\beta +1) (\lambda +1) x_{3}^2-631\big)\big)-560 (\beta +1) (\lambda +1) x_{2}^4\\
  &+x_{2}^2 \big(378 (\beta -1) \lambda ^2+(2081 \beta +191) \lambda+1811 \beta-910 (\beta +1) (\lambda +1) x_{3}^2+677\big)+126 (\beta -1) \lambda ^2 \big(5 x_{3}^2-8\big)\\
  &-x_{3}^2 \big(-1317 \beta +350 (\beta +1) x_{3}^2+573\big)-\lambda  \big(2151 \beta+350 (\beta +1) x_{3}^4+3 (509-541 \beta ) x_{3}^2\big)\big)\bigg) \Bigg/ (2520 (\lambda +1)^3),
    \end{aligned}
\end{equation}
\begin{equation}\label{eq26}
    \begin{aligned}
    \hat{u}^{(1)}_{3}&=   \bigg((\beta +1) (19 \lambda +16) x_{3} \big(-35 (\beta +1) (\lambda +1) x_{1}^4-x_{1}^2 \big(-3 \big(42 (\beta -1) \lambda ^2+(281 \beta +71) \lambda +167 \beta +41\big)\\
    &+1050 (\beta +1) (\lambda +1) x_{2}^2-35 (\beta +1) (\lambda +1) x_{3}^2\big)-35 (\beta +1) (\lambda +1) x_{2}^4+x_{2}^2 \big(3 \lambda  (71 \beta -42 (\beta -1) \lambda )\\
    &+123 \beta +843 \lambda +35 (\beta +1) (\lambda +1) x_{3}^2+501\big)+2 (\beta +1)\big(x_{3}^2-1\big) \big(27 (\lambda -1)+35 (\lambda +1) x_{3}^2\big)\big)\bigg)\Bigg/ (1260 (\lambda +1)^3).  
    \end{aligned}
\end{equation}
 The $O(1)$ velocity field above, for planar extension, exhibits a simple $(1+\lambda)^{-1}$\!-scaling\,(equations (\ref{eq21}-\ref{eq23}) with $\beta=1$), allowing for the $\lambda$-dependence to be incorporated into a re-scaled Peclet number. In contrast, the $O(Ca)$ correction exhibits a complicated non-factorizable dependence on $\lambda$ (see equations (\ref{eq24}-\ref{eq26})), which leads, in principle, to a non-trivial dependence of the finite-$Ca$ streamline topology on $\lambda$.
 
\subsection{Langevin simulations}\label{appA2}
The mass transport for the interior problem of interest is governed by the convection-diffusion equation for the scalar concentration field, with convection being driven by the Stokesian velocity field to $O(Ca)$: 
	\begin{equation}\label{eq27}
		\frac{\partial c}{\partial \tau}+ Pe \left( \bm{\hat{u}}^{(0)}(\bm{x})+Ca\, \bm{\hat{u}}^{(1)}(\bm{x}) \right)
  \cdot \bm{\nabla}c= \nabla^{2} c.
	\end{equation}
In (\ref{eq27}), velocity and time are scaled by $a \dot{\gamma}$ and $a^2 /D$, respectively, $D$ being the diffusivity coefficient; $\bm{\hat{u}}^{(0)}$ and $\bm{\hat{u}}^{(1)}$ are defined by (\ref{eq21}-\ref{eq23}) and (\ref{eq24}-\ref{eq26}). The Peclet number is defined as $Pe=\dot{\gamma} a^2 / D$, with $Pe\gg 1$ denoting the convection dominant limit. Equation (\ref{eq27}) is solved starting from an initially uniform distribution and with the drop acting as a perfectly absorbing boundary. The suitably normalized concentration $c$ satisfies $c=1$ at $\tau=0$ for $|\bm{x}|\le 1$, and $c=0$ at the surface of the deformed drop as defined above

The interior streamlines, for the planar linear flows under consideration, do not conform to any obvious symmetries, precluding the use of standard coordinate systems. Therefore, we use Langevin simulations to simulate the transport process that, on average, is described by (\ref{eq27}). The Langevin equations \cite{gardiner2004} for the individual tracers are:
\begin{equation}\label{eq28}
	d\bm{x}= Pe\left( \bm{\hat{u}}^{(0)}(\bm{x})+Ca\,\bm{\hat{u}}^{(1)}(\bm{x}) \right) d\tau+ \sqrt{2}~ d\bm{W}(\tau).
\end{equation}
Here, $\hat{\bm{u}}(\bm{x})$ convects the tracer particle at position $\bm{x}(\tau)$, and the final term, $d\bm{W}(\tau)$, corresponds to the standard Wiener process satisfying $<d\bm{W}(\tau)>=0$, $<(d\bm{W})^2>=d \tau$. We solve the Langevin equations using a $4^{th}$ order Runge-Kutta technique with a sufficiently small time step that satisfies $Pe\,d\tau\ll 1$, ensuring numerical convergence. In the numerical implementation $\displaystyle d\bm{W}(\tau) = \sqrt{d\tau} N(0,1)$, $N(0,1)$ being a Gaussian random variable with zero mean and unit variance. For a sufficiently large initial number of tracers, the ensemble average of the tracer concentration converges to the solution of the convection-diffusion equation. In all our runs, we start from $10^5$ tracers uniformly distributed within the drop, with the absorbing boundary condition implemented at each time step by removing any tracers that end up crossing the drop surface. To address the decrease in the number of tracers with time, and the associated increase in the amplitude of the tracer number fluctuations, we implemented the following replenishment protocol: when the tracers reduce to half the initial number, an equal number of new tracers are introduced at randomly chosen locations. This ensures that the tracer number never dips below $50000$ during the course of any simulation, in turn ensuring accurate long-time averages required for the $Nu$ calculation\,(see below).
 
 Using the time dependent volume-averaged concentration obtained from the simulations, we calculate the non-dimensional transport rate as: $\displaystyle Nu=\frac{2}{\bar{c}} F$\cite{kronig1951,juncu2010}. Here, $\displaystyle\bar{c}(\tau)= \frac{1}{V}\int\!\!c(\bm{x},\tau)d\rm{V}$ is the volume-averaged concentration, and $\displaystyle F=-\frac{1}{4\pi}\oint_{S} \bm{n} \cdot \nabla c(\bm{x},\tau)d\rm{S}$ is the surface averaged flux, with the area of the undeformed drop\,($4\pi$) used for normalization.  Using the divergence theorem, $\displaystyle F=-\frac{1}{4\pi} \oint\nabla^2 c(\bm{x},\tau)d\rm{V}$ and further use of the convection-diffusion equation reduces this to $\displaystyle F=-\frac{1}{4\pi} \oint \frac{\partial c}{\partial \tau}(\bm{x},\tau)d\rm{V}$. Since the drop attains a steady non-spherical shape with volume equal to the original undeformed drop\,($\frac{4\pi}{3}$), one may write $\displaystyle F=-\frac{1}{3} \frac{\partial}{\partial \tau} \left[ \frac{1}{V} c(\bm{x},\tau) d\rm{V} \right] =-\frac{1}{3} \frac{\partial \bar{c}}{\partial \tau} $. Hence, one finally obtains $\displaystyle Nu=-\frac{2}{3} \frac{1}{\bar{c}} \frac{\partial \bar{c} }{\partial \tau}~ \text{or}~ -\frac{2}{3} \frac{\partial \log \bar{c}}{\partial \tau}$. In each of our simulations, $Nu$ exhibits an algebraic decrease proportional to $t^{-\frac{1}{2}}$ for short times, eventually asymptoting to a plateau with superposed small amplitude fluctuations. The $Nu$-values plotted in the main manuscript correspond to time independent values obtained by taking a decade-long time average in this long-time plateau regime.

\begin{figure}
		\centering
		\includegraphics[scale=0.4]{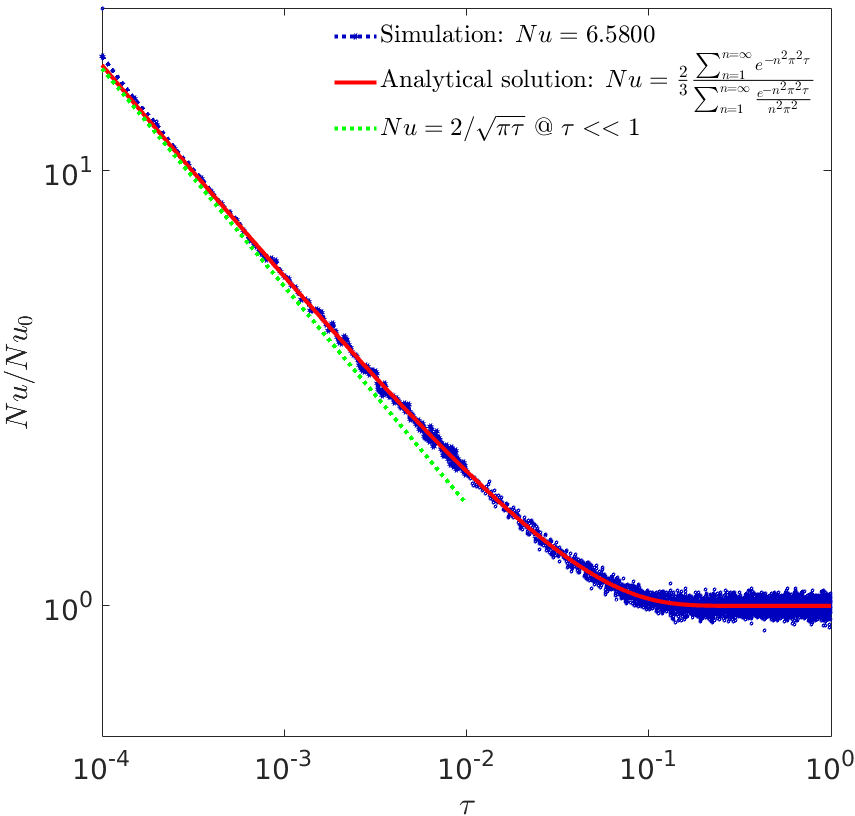}
		\caption{Comparison of the $Nu$ obtained from the Langevin simulations, for $Pe = 0$ with the exact analytical solution defined in the text, and the short-time asymptote.}\label{fig3l} 
	\end{figure}

To validate our simulations, we consider the analytically soluble pure diffusion problem corresponding to $Pe=0$. For this case, the time-dependent concentration field is given by $\displaystyle c(\bm{x},\tau)=\frac{2}{\pi r} \sum_{n=1}^{\infty} \frac{(-1)^{n+1}}{n} \sin(n\pi r) e^{-n^{2}\pi^{2} \tau}$, with the volume-averaged concentration given by 
$\displaystyle \bar{c}(\tau)=6 \sum_{n=1}^{\infty} \frac{e^{-n^2 \pi^2 \tau}}{n^2 \pi^2}$; accordingly, $Nu_{0}=\frac{2}{3\pi^2}$, the subscript `0' denoting the diffusion-limit. For sufficiently short times, the diffusive transport occurs across a thin $O(\tau^\frac{1}{2})$ concentration boundary layer just beneath the drop surface that may now be approximated as an infinite plane; this yields $Nu  \approx 2/\sqrt{\pi \tau}$ for $\tau \ll 1$. Figure \ref{fig3l} shows the $Nu$ obtained from the simulations alongside both the exact analytical solution and the short-time asymptote above, and highlights the excellent agreement between them. It is worth noting that the aforementioned short-time asymptote continues to be valid for all $Pe$, although for large $Pe$, it gives way to convection-induced oscillations for $\tau \gtrsim O(Pe^{-1})$. 

\subsection{Boundary element method}\label{BEM}	
	The boundary integral formulation for the Stokes equations expresses the velocity field at any point $\bm{x}_{0}$ in terms of distributions of point forces\,(the single-layer potential) and point-force dipoles/sources\,(the double-layer potential) on the surface of the drop, and may be written in non-dimensional form as \cite{kennedy1994,pozrikidis1992, pozrikidis2002}: 
	\begin{equation}\label{eq22a}
		\begin{aligned}
			\bm{u}_{j}(\bm{x_{0}})&=\frac{Ca}{\lambda} u_{j}^{\infty}(\bm{x_{0}}) -\frac{1}{8 \pi \lambda} \int_{S} \Delta f_{i}(\bm{x})  G_{ij}(\bm{x},\bm{x_{0}}) dS(\bm{x})+ \frac{1-\lambda}{8 \pi \lambda} \int_{S} \bm{u}_{i}(\bm{x}) T_{ijk}(\bm{x}, \bm{x_{0}}) \bm{n}_{k}(\bm{x}) dS(\bm{x}).
		\end{aligned}
	\end{equation} 
	 In the above relation, $S$ is the surface of the deformed drop, and $\bm{n}$ is the unit normal vector pointing into the ambient fluid. The first term on the RHS corresponds to the imposed flow $\bm{u}^{\infty}(\bm{x})= \bm{\Gamma} \cdot \bm{x}$ with $\bm{\Gamma}$ defined in section \ref{appA1}. In the second term, $\Delta f_{i}(\bm{x})=(\sigma_{ij}-\lambda \hat{\sigma}_{ij}) \cdot \bm{n}_{j}$ is the discontinuity in the interfacial force density, while the kernels of the single and double layer potentials, $G$ and $T$, are the Stokeslet and its corresponding stress tensor, given by $\displaystyle 	G_{ij}(\bm{x,x_{0}})=\frac{\delta_{ij}}{|\bm{\hat{x}}|}+\frac{\hat{x}_{i} \hat{x}_{j}}{|\bm{\hat{x}}|^{3}}$ and $\displaystyle T_{ijk}(\bm{x,x_{0}})=-6\frac{\hat{x}_{i} \hat{x}_{j} \hat{x}_{k}}{|\hat{\bm{x}}|^{5}}$, 	where $\hat{\bm{x}}=\bm{x-x_{0}}$. Taking the limit of $\bm{x}_{0}$ approaching the surface of the drop, one obtains in the general case a Fredholm integral equation of the second kind for the interfacial velocity: 
 \begin{equation}\label{eq24a}
		\begin{aligned}
\bm{u}_{j}(\bm{x_{0}})&=\frac{2 Ca }{1+\lambda} u_{j}^{\infty}(\bm{x_{0}}) -\frac{1}{4 \pi (1+\lambda)} \int_{S} \Delta f_{i}(\bm{x}) G_{ij}(\bm{x},\bm{x_{0}}) dS(\bm{x})+ \frac{1-\lambda}{4 \pi (1+\lambda)} \int_{S}^{PV} \bm{u}_{i}(\bm{x}) T_{ijk}(\bm{x}, \bm{x_{0}}) \bm{n}_{k}(\bm{x}) dS(\bm{x}),
		\end{aligned}
	\end{equation}
 where $PV$ signifies the improper-integral that results from the original double layer potential when $\bm{x}_{0}$ is located on the $S$. For $\lambda=1$, (\ref{eq24a}) reduces to the following much simpler explicit equation for the velocity field:
\begin{equation}\label{eq24b}
		\begin{aligned}
\bm{u}_{j}(\bm{x_{0}})&=Ca~ u_{j}^{\infty}(\bm{x_{0}}) -\frac{1}{8 \pi } \int_{S} \Delta f_{i}(\bm{x}) G_{ij}(\bm{x},\bm{x_{0}}) dS(\bm{x}).
		\end{aligned}
	\end{equation}
 
We obtain the interior velocity field by solving either (\ref{eq24a}) or (\ref{eq24b})  for a deformed drop, and then use it as an input for the Langevin simulations to obtain $Nu$. The evaluation of the integrals required for this purpose is done along the same lines as Kennedy and Pozrikidis \cite{pozrikidis1992,pozrikidis2002,kennedy1994}. Thus, we divide the drop surface into triangular elements with each element being assigned an array of six marker points\,(nodes), this leading to a system of linear equations for the nodal velocity fields, to be solved at every time step, for $\lambda \neq 1$. The solution\,(velocity field) obtained at the nodes is used, via subsequent interpolation, to evaluate the surface integral over a given triangular element, and thence the interior velocity field\,(required for plotting the streamlines). 

\begin{figure}
\centering
\includegraphics[scale=0.35]{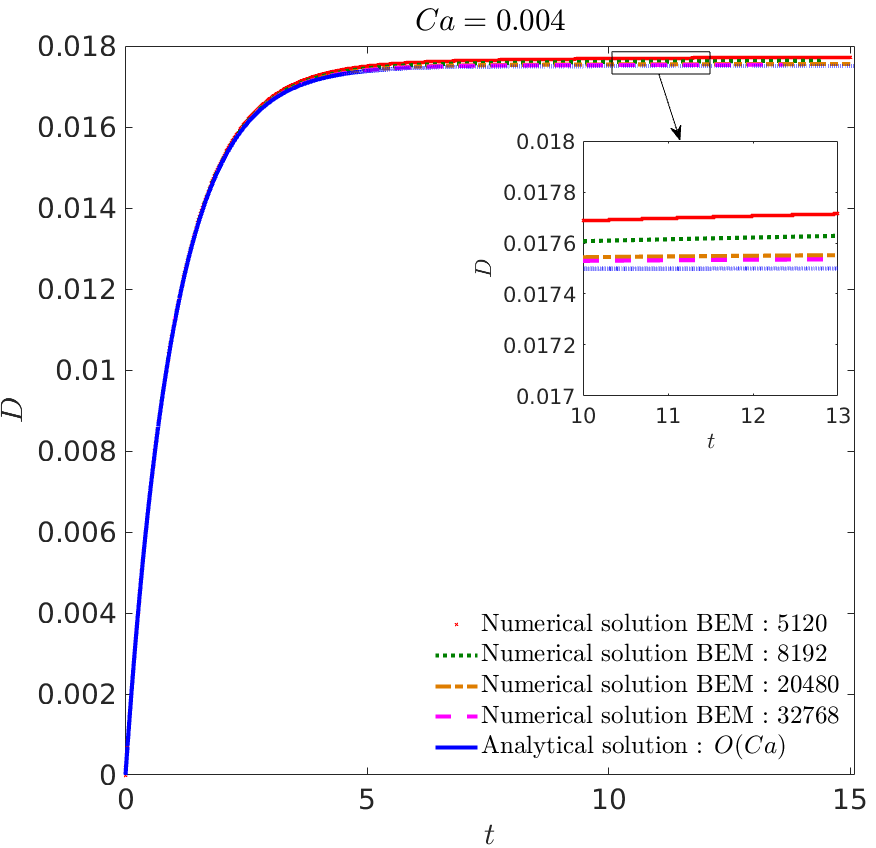}
\caption{Taylor deformation parameter $D$ plotted as function of $t$ for planar extension for $Ca=0.004$, $\lambda=1$; the number of elements for the $BEM$-curves correspond to $5120$, $8192$, $20480$, and $32768$; the solid blue line is the $O(Ca)$ given by $D =2b_{1}^{(1)}Ca$.}		\label{fig9la}
	\end{figure}

\begin{figure}
\centering
\includegraphics[scale=0.35]{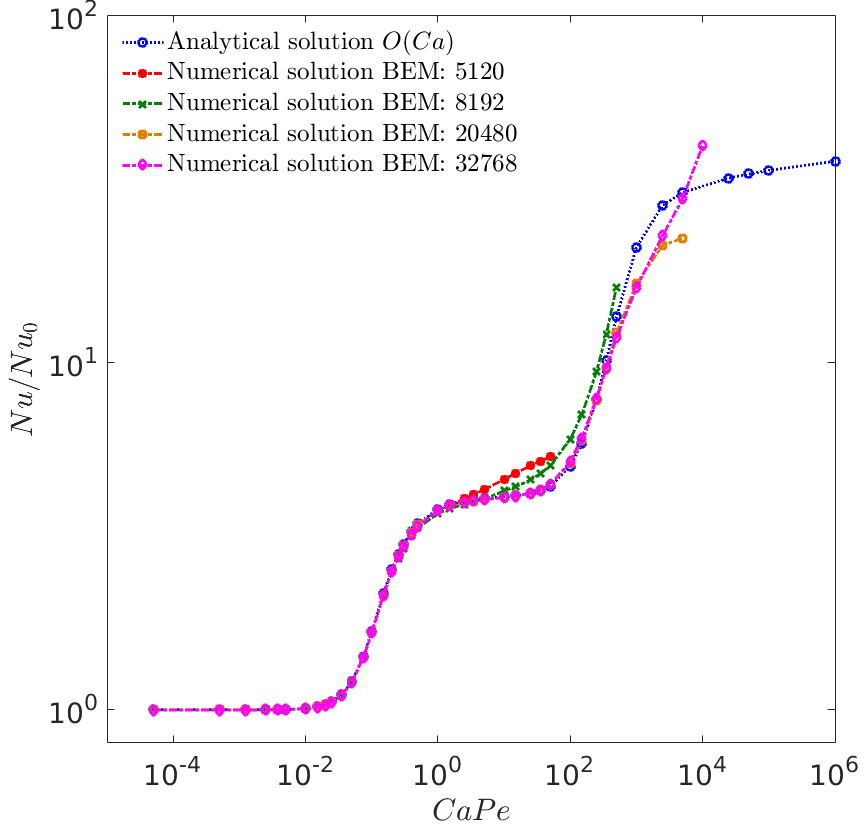}
\caption{Comparison of $Nu/Nu_{0}$ plotted as a function of $PeCa$ for $Ca=0.005$, $\lambda=1$, for planar extension; the BEM-curves corresponding to $5120,~ 8192,~20480$, and $32768$ elements are compared with the $Nu$-curve based on the $O(Ca)$ velocity field.}		\label{fig9l}
	\end{figure}

It is worth mentioning that the streamline topology for small but finite $Ca$ is particularly sensitive to surface discretization\,(the number of surface elements), requiring sufficiently high resolution for a given accuracy. The lack of resolution is very likely the reason the deformation-induced change in streamline topology went unnoticed in the earlier effort of \cite{kennedy1994}. We therefore, study the grid-independence of our BEM simulations via a comparison of both the macroscopic drop deformation parameter, and the actual interior streamlines, with those obtained from the small-$Ca$ theory. Fig\,\ref{fig9la} shows the Taylor deformation parameter $D=(L-B)/(L+B)$\,($L$ and $B$ denoting the major and minor axes of the ellipsoidal drop in the plane of the flow) as a function of time, for a varying number of elements\,($5120$, $8192$, $20480$ and $32768$), along with $O(Ca)$ prediction obtained using (\ref{eq14}) for a planar extensional flow. The BEM results begin to converge only for $20480$ and $32768$ elements. We also compared the $Nu-Pe$ curves for various gird sizes. 
Fig\,\ref{fig9l} compares the scaled $Nu$-curves based on the analytical $O(Ca)$ velocity field, and those obtained from the BEM simulations, for $Ca = 0.005$. Again, it is only for $20480$ and $32768$ surface elements that one begins to get a good agreement for the secondary rise, that begins for finite values of $PeCa$, and is driven by the spiralling-streamline topology. Note that even for these many elements, the BEM $Nu$-curve departs from the analytical one prior to the secondary diffusion-limited plateau. Accurately capturing the secondary plateau requires the BEM simulations to faithfully capture the complete invariant tori that exist for the said $Ca$\,(=0.005), and this appears to require even higher resolution than that used.   

\begin{figure}
		\centering
\includegraphics[scale=0.175]{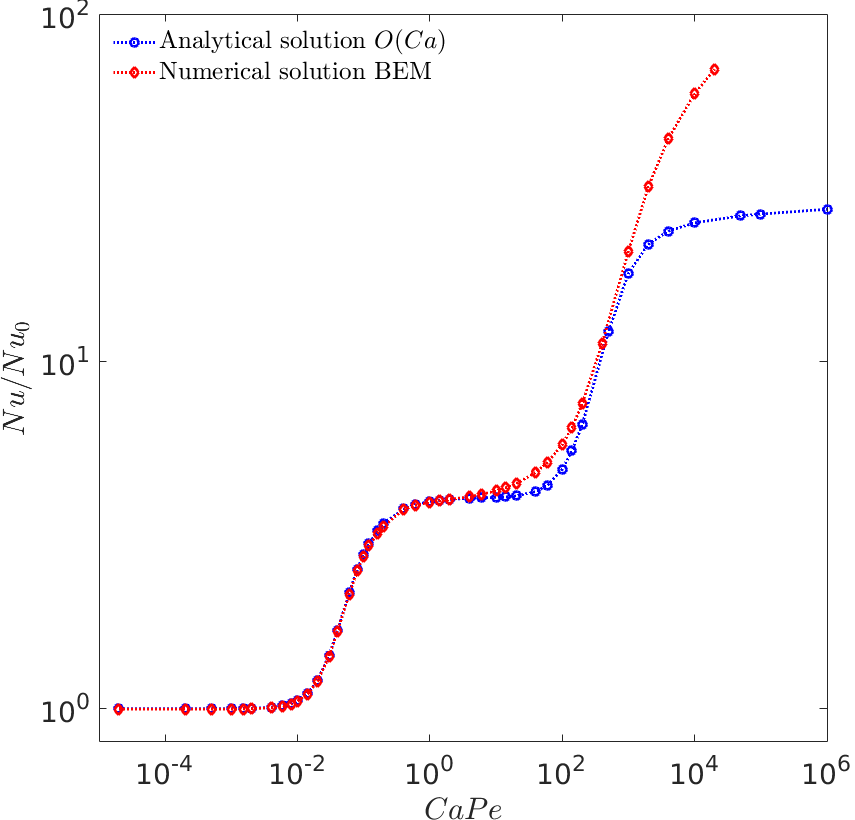} \quad
\includegraphics[scale=0.175]{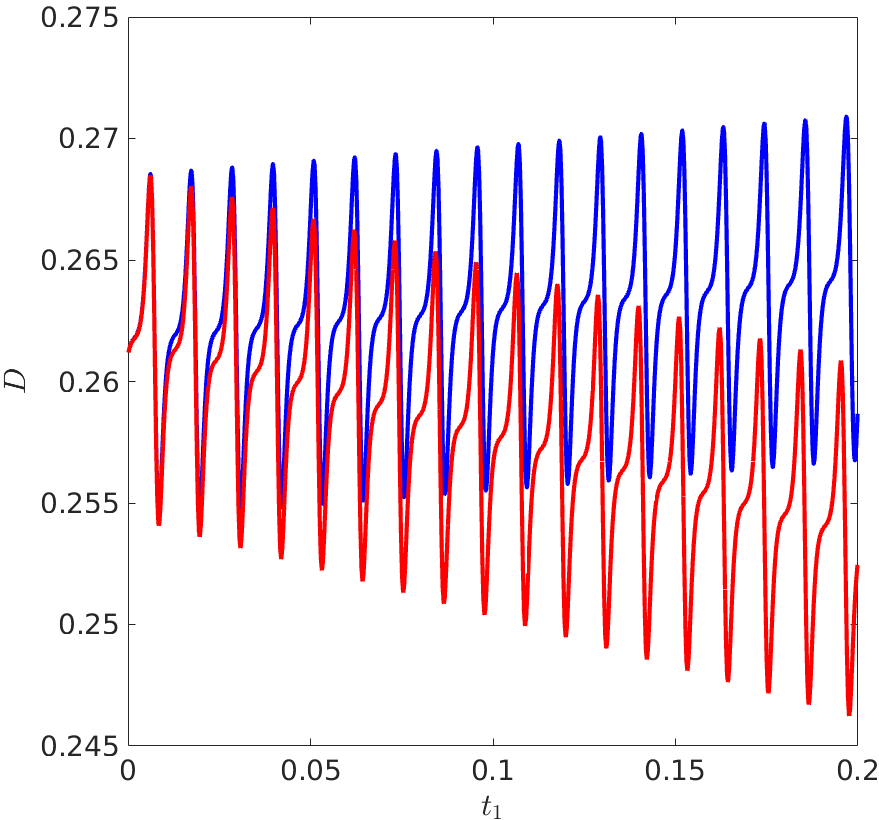}\quad
\includegraphics[scale=0.175]{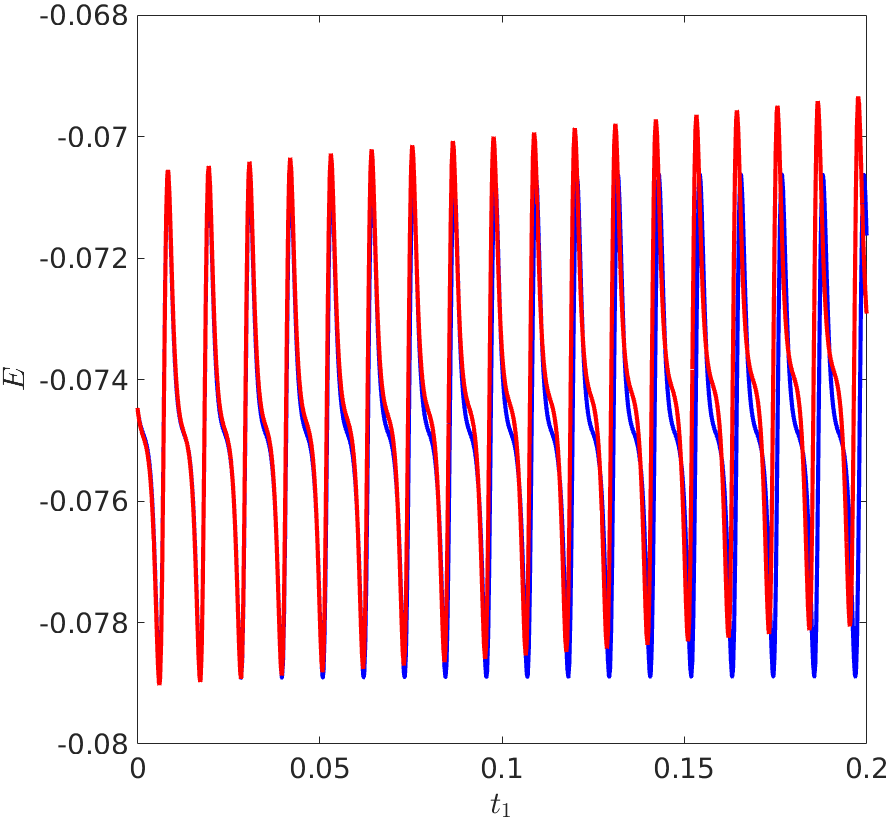} \\
(a) \hspace{3.5cm} (b) \hspace{3.5cm} (c)\\
\includegraphics[scale=0.175]{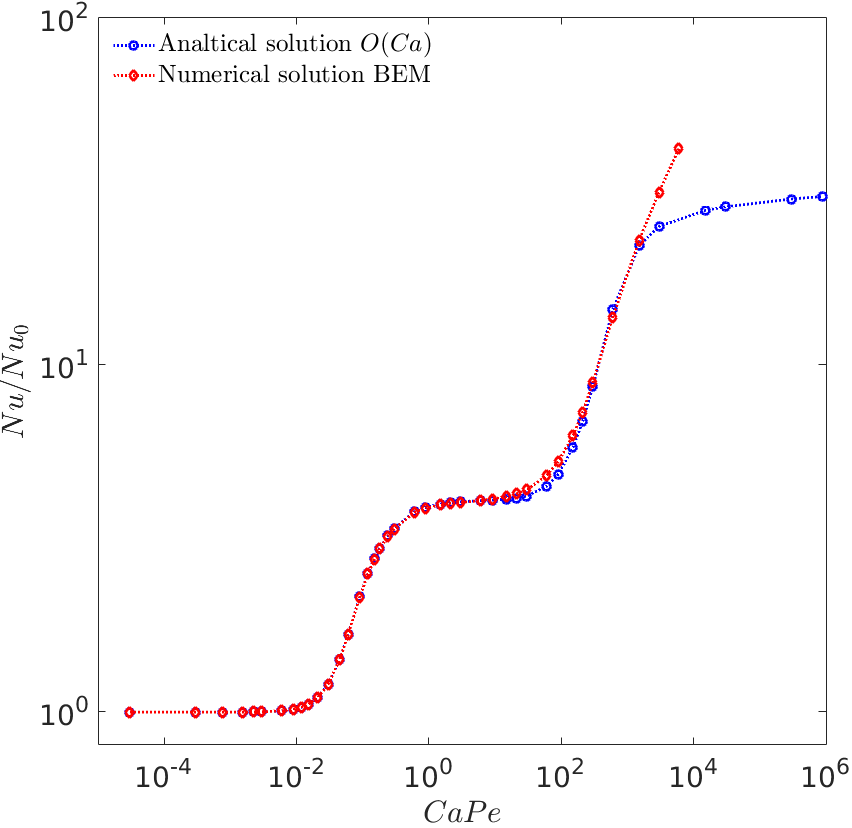} \quad
\includegraphics[scale=0.175]{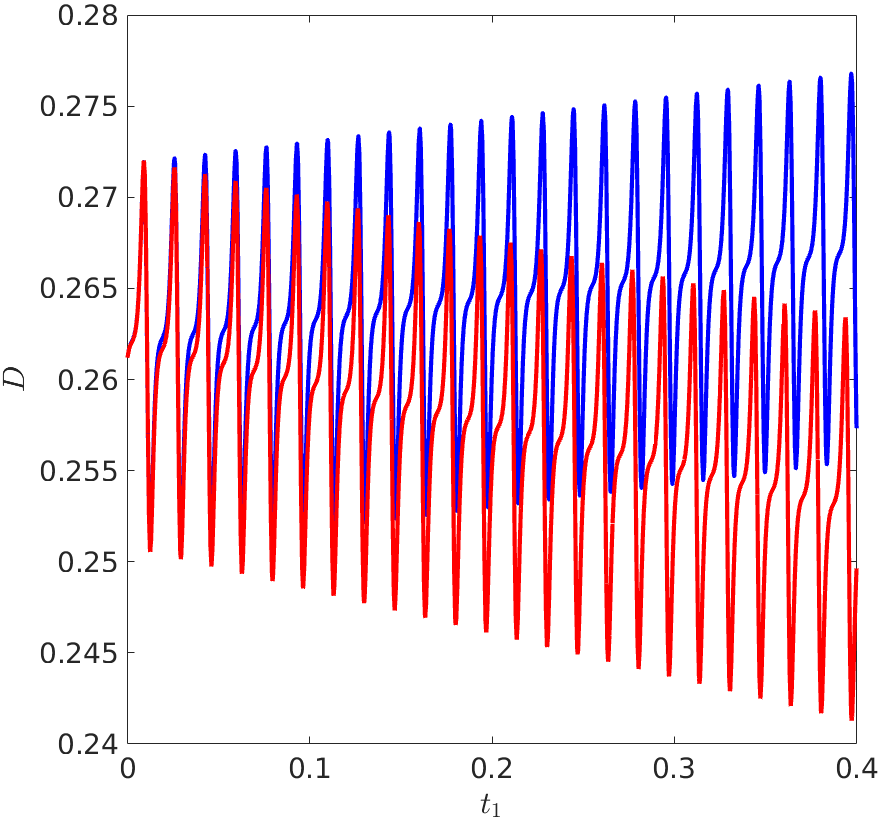}\quad
\includegraphics[scale=0.175]{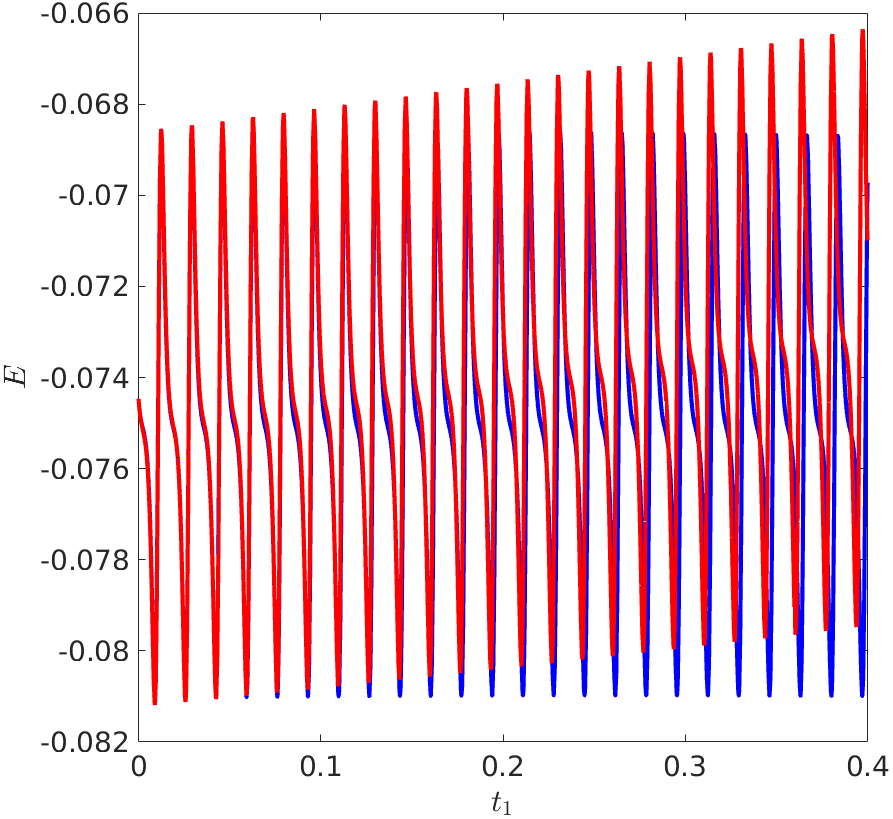} \\
(d) \hspace{3.5cm} (e) \hspace{3.5cm} (f)\\
\includegraphics[scale=0.175]{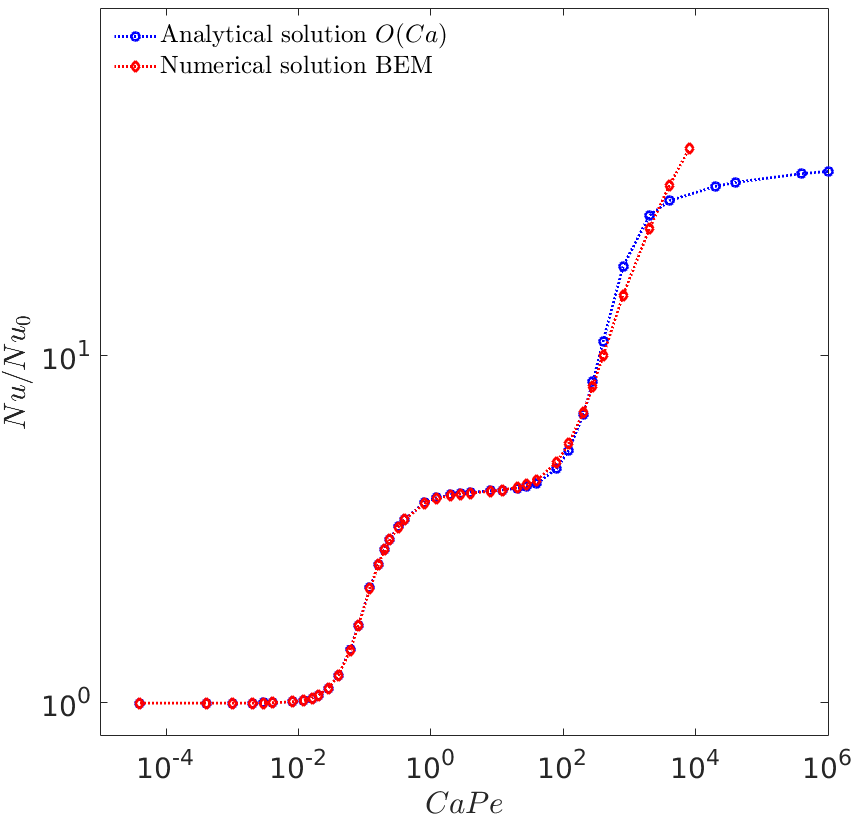} \quad
\includegraphics[scale=0.175]{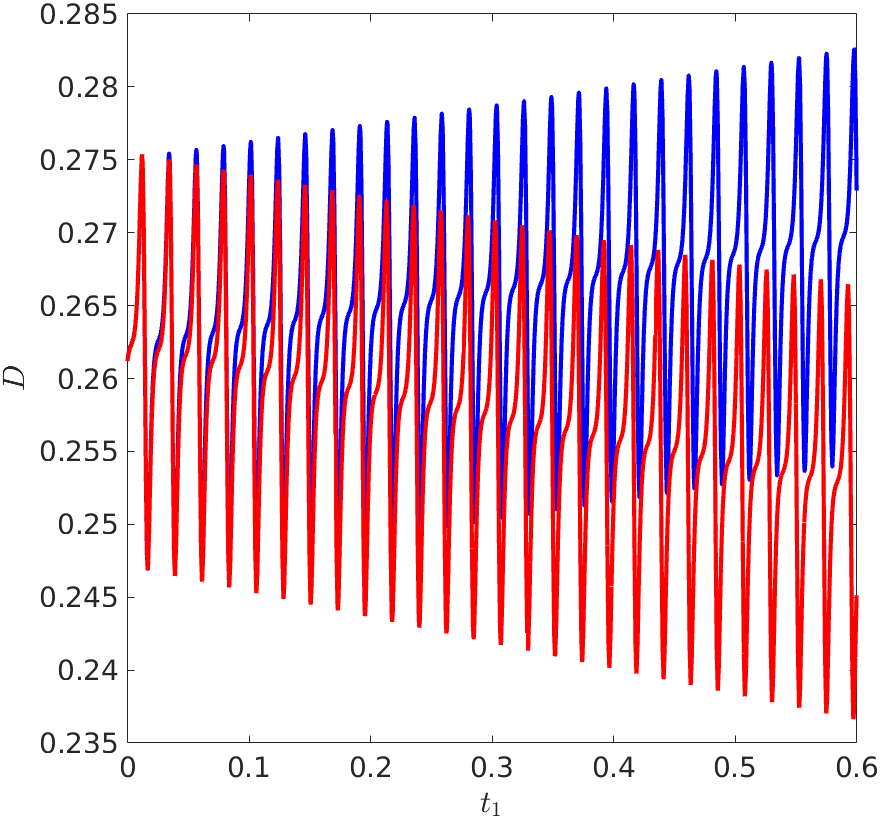}\quad
\includegraphics[scale=0.175]{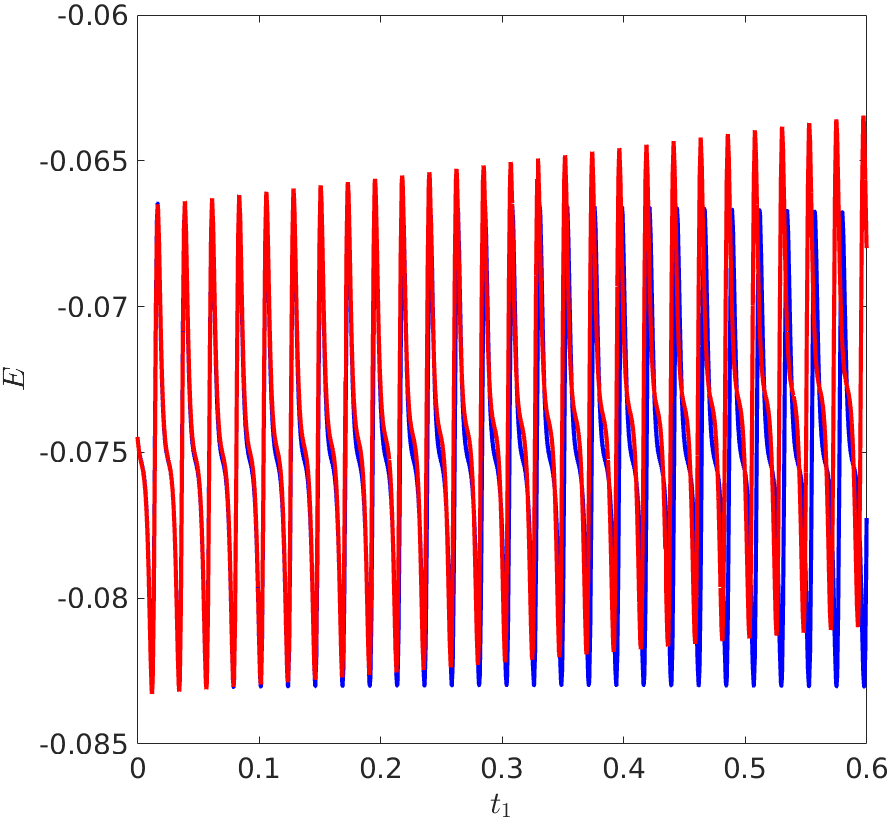} \\
(g) \hspace{3.5cm} (h) \hspace{3.5cm} (i)\\
\includegraphics[scale=0.175]{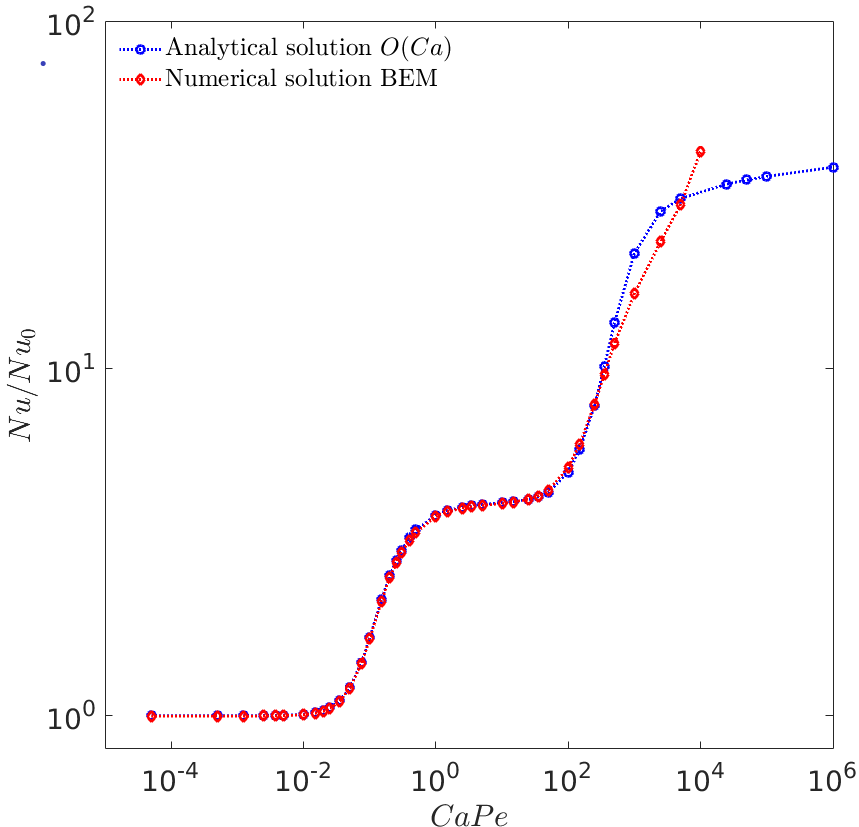} \quad
\includegraphics[scale=0.175]{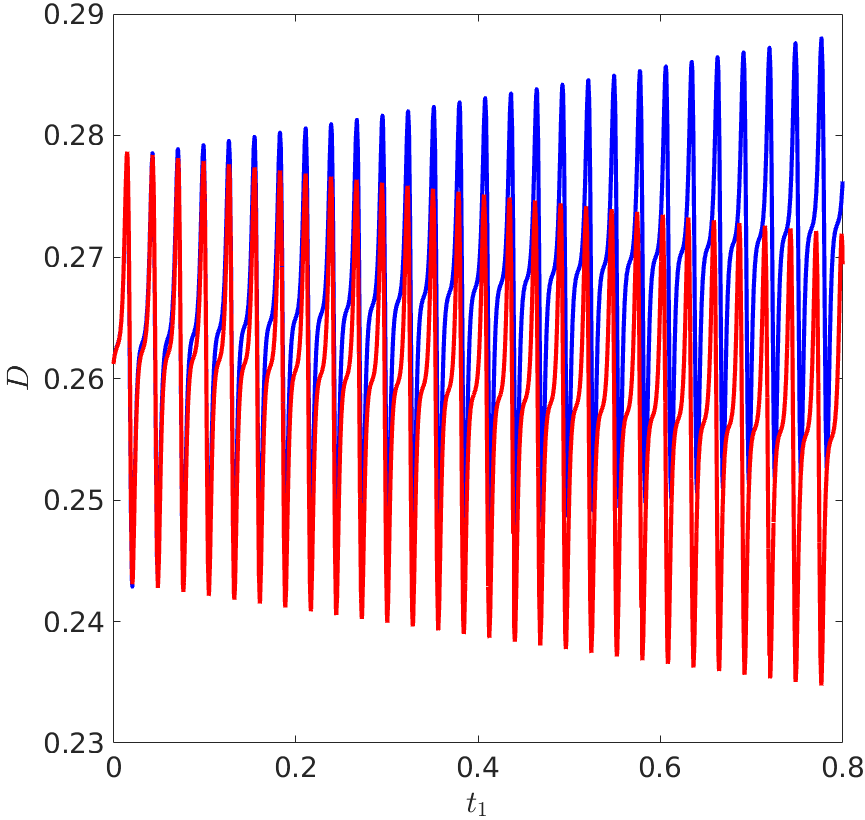}\quad
\includegraphics[scale=0.175]{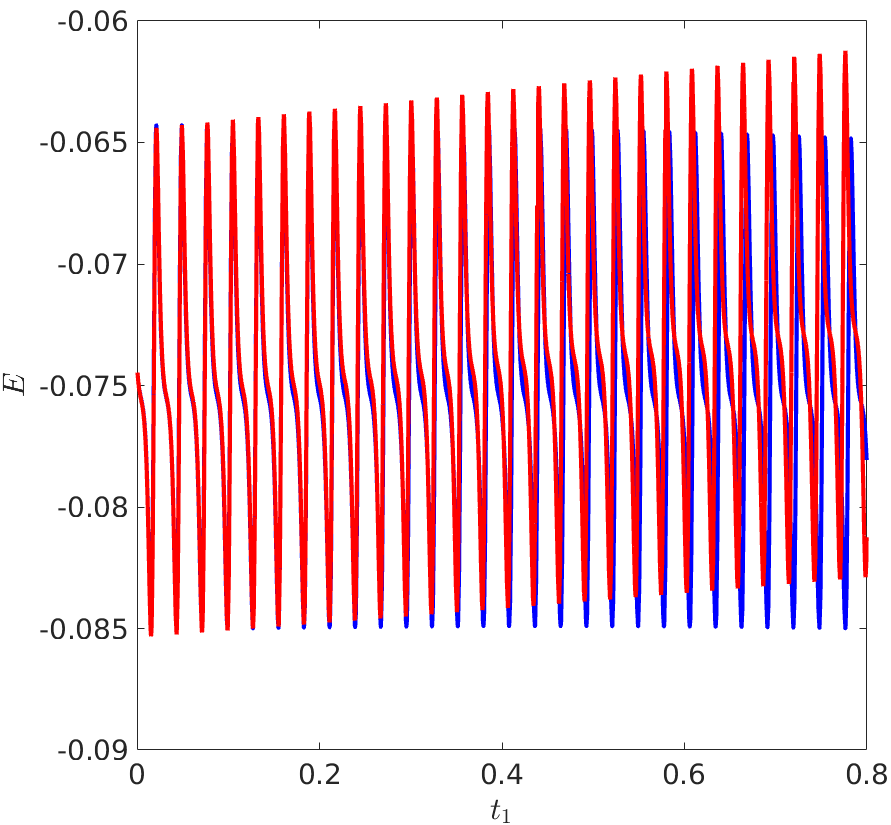} \\
(j) \hspace{3.5cm} (k) \hspace{3.5cm} (l)\\
\includegraphics[scale=0.175]{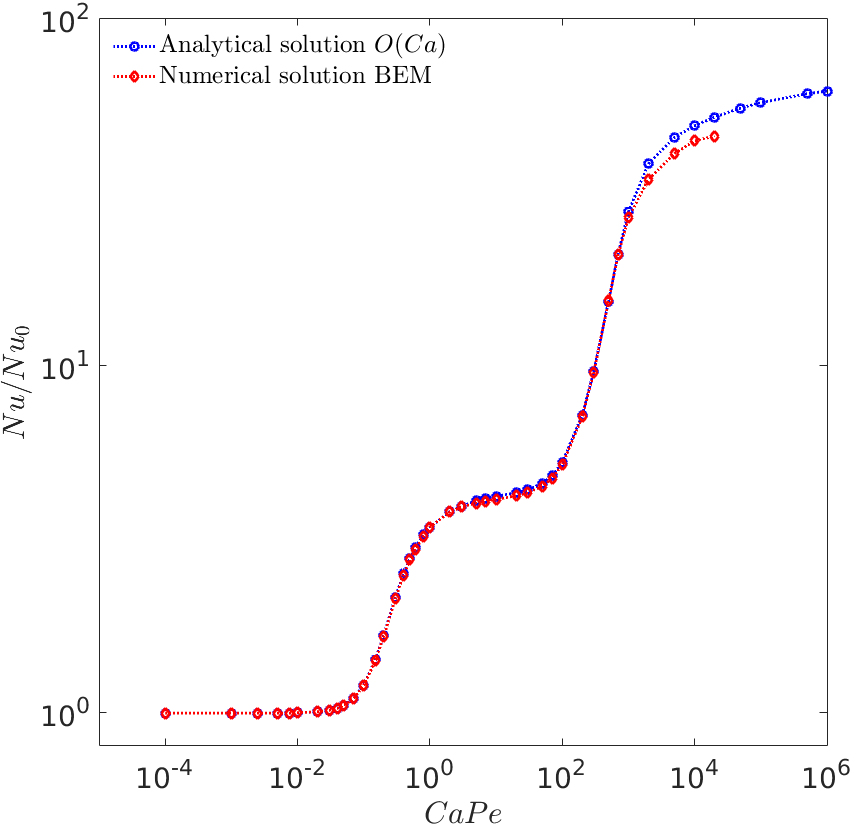} \quad
\includegraphics[scale=0.175]{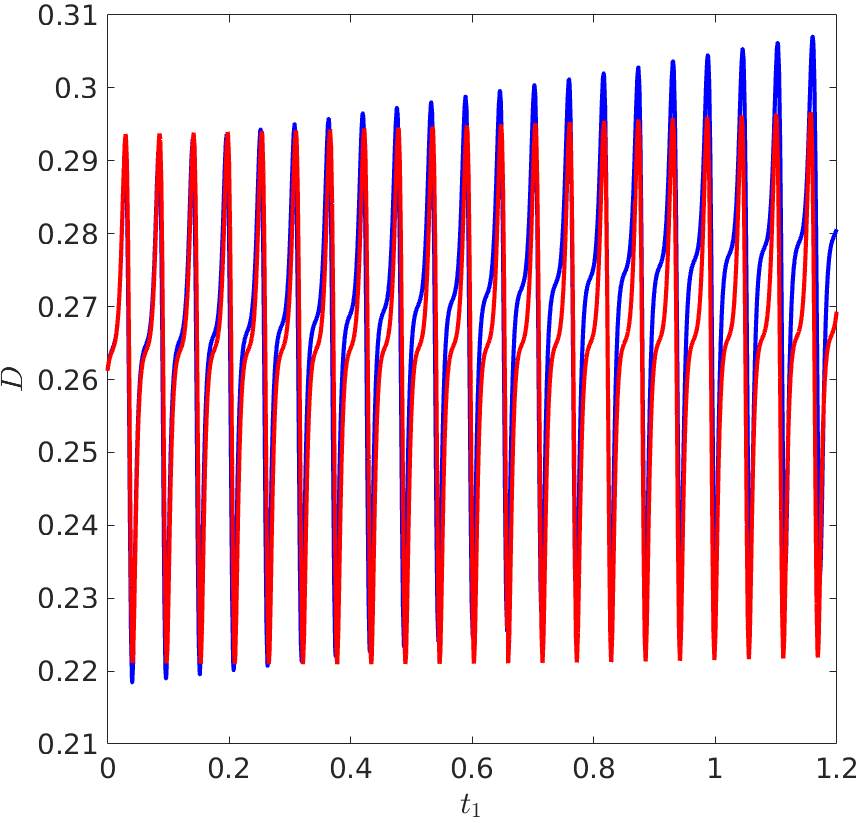}\quad
\includegraphics[scale=0.175]{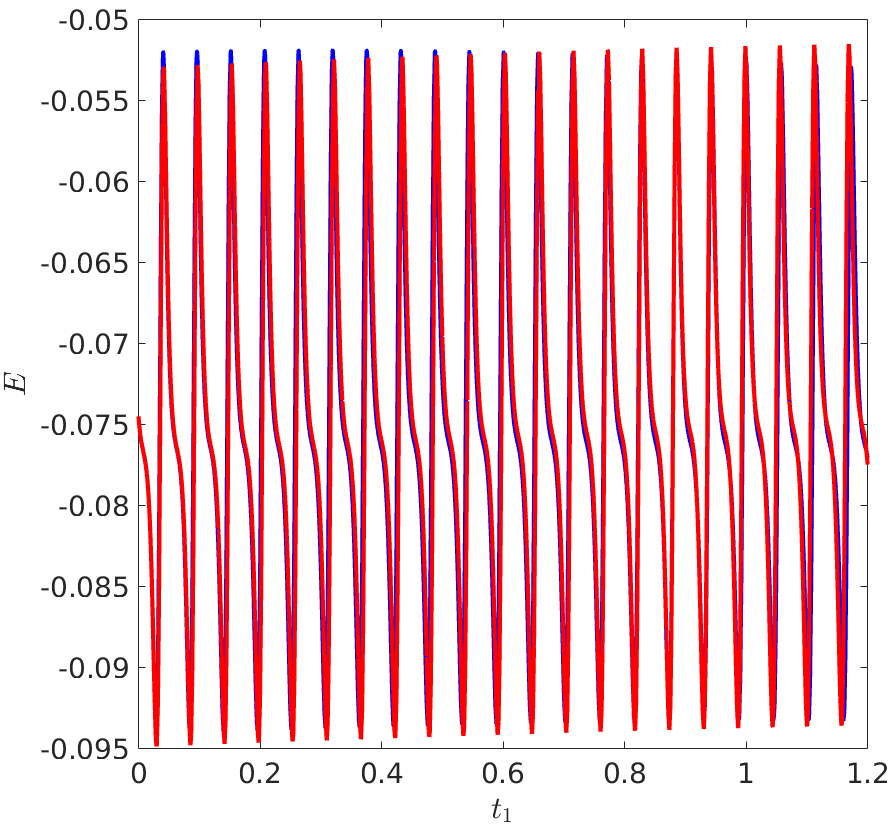} \\
(m) \hspace{3.5cm} (n) \hspace{3.5cm} (o)\\
\caption{Scaled $Nu/Nu_{0}$-curves plotted as a function of $CaPe$ for various $Ca$ values and $\lambda=1$ with $32768$ grid element results, with the streamline topology as $D$ vs $\tau$ and $E$ vs $\tau$. Here, we compare the numerical results (solid red line) with the results obtained by the analytical solution (solid blue line);
 $Ca=0.002$:(a)-(c), $Ca=0.003$:(d)-(f), $Ca=0.004$:(g)-(i), $Ca=0.005$:(j)-(l), $Ca=0.01$:(m)-(0)}	\label{fig10l}
	\end{figure}

Figs\,\ref{fig10l} show a comparison of the scaled $Nu$-curves obtained from the BEM simulations, for $Ca$ increasing from $0.002$ to $0.01$, with the one obtained using the $O(Ca)$ velocity field. Rather counter-intuitively, the agreement of the aforementioned secondary rise exhibited by the BEM curves, with that of the $O(Ca)$-field-based $Nu$-curve, improves with increasing $Ca$. This is because accurate capture of the secondary rise\,(and the eventual plateau) requires the BEM simulations to capture the smallest length associated with the finite-$Ca$ flow field - the pitch of the spiralling streamlines for small $Ca$. For a given number of surface elements, there is a threshold $Ca$ below which the aforementioned pitch reduces below the average size of a surface element, and this leads to a progressive deterioration in the accuracy for smaller $Ca$. This feature leads to the best agreement between the BEM and the $O(Ca)$ $Nu$-curves occurring at a small but finite $Ca$; for the case shown, this is $Ca = 0.01$. Agreement at smaller $Ca$ is limited by drop-surface discretization, while that at larger $Ca$ is limited by the range of validity of the small-$Ca$ theory. The accompanying streamline plots in the figures above, that show the adiabatic invariants\,($D$ and $E$) as a function of the slow time scale\,($t_1 = Ca\,t)$, clearly show that the improving $Nu$-comparison is due to the improving agreement between the numerical and analytical streamlines with increasing $Ca$. However, even for the highest $Ca$, the secondary plateau regime remains numerically inaccessible owing to the BEM-streamline departing from the analytical one for $t_1 \gtrsim O(1)$.  
 
\bibliographystyle{apsrev4-1}
\bibliography{supply_prl}
